\renewcommand{\v}[1]{\ensuremath{\mathbf{#1}}} 
\newcommand{\gv}[1]{\ensuremath{\mbox{\boldmath$ #1 $}}} 
\newcommand{\ten}[1]{\ensuremath{\tensorsym{#1}}}
\newcommand{\uv}[1]{\ensuremath{\mathbf{\hat{#1}}}} 
\newcommand{\abs}[1]{\left| #1 \right|} 
 
\newcommand{\pdd}[2]{\frac{\partial^2 #1}{\partial #2^2}} 
\newcommand{\grad}[1]{\gv{\nabla} #1} 
\renewcommand{\div}[1]{\gv{\nabla} \cdot #1} 
\let\baraccent=\= 
\renewcommand{\=}[1]{\stackrel{#1}{=}} 
\DeclareMathAlphabet{\mathsfsl}{OT1}{cmss}{m}{sl}

\documentclass[%
 preprint,
 amsmath,amssymb,
 aps,
]{revtex4-2}

\usepackage{graphicx}
\usepackage{dcolumn}
\usepackage{bm}

\usepackage{tensor} 
\usepackage{isomath} 
\usepackage{mathrsfs} 

\begin{document}

\preprint{Draft for External Revision}

\title{Stokesian Dynamics with odd viscosity}

\author{Hang Yuan}
\affiliation{%
 Applied Physics Graduate Program, Northwestern University, Evanston, IL 60208
}%

\author{Monica Olvera de la Cruz}
 \email{m-olvera@northwestern.edu}
\affiliation{
Department of Materials Science and Engineering, Northwestern University, Evanston, IL 60208
}%
\affiliation{
Department of Chemistry, Northwestern University, Evanston, IL 60208
}%
\affiliation{
Department of Physics and Astronomy, Northwestern University, Evanston, IL 60208
}%

\date{\today}

\begin{abstract}
Stokesian Dynamics is a well-established computational method for simulating dynamics of many particles suspended in a conventional passive fluid medium. Active fluids composed of self-propelling particles with broken time reversal symmetry permit the emergence of a so-called odd viscosity. In this work, we extended the conventional Stokesian Dynamics formalism to incorporate the additional hydrodynamic effects due to odd viscosity, which enables simulating collective behaviors of many particles suspended in an active fluid medium with both even viscosity and odd viscosity.
\end{abstract}

\maketitle


\section{Introduction}

Understanding and predicting the macroscopic properties, such as the diffusion coefficient\cite{Dhont1996}, thermal conductivity and rheology\cite{Mewis2011}, of a suspension of particles from its micro-structural mechanics is a long-standing science and engineering problem. In order to address this problem, a vast number of different simulation techniques have been developed in past decades \cite{PhysRevE.52.1734,Pozrikidis2002,Keaveny2014,Krueger2016,Fiore2019,Bolintineanu2014}. Stokesian Dynamics\cite{Brady1988} is a well-known particle-based approach for dynamically simulating the collective behaviors of many particles suspended in a fluid medium. It has been successfully applied to a diverse range of particle suspensions for understanding phenomena such as sedimentation\cite{Revay1992,Cichocki1993,Nakayama2005}, phase transition of colloidal gels\cite{Dickinson2013,Varga2015,Landrum2016} and discontinuous shear thickening\cite{PhysRevLett.111.218301,Mari2015,PhysRevLett.115.228304}, to name a few.

Hydrodynamic interactions have played an important role in describing suspensions of particles in viscous fluids for decades. In the 1990s, Aron and co-workers\cite{Avron1995,avron1998odd} showed that the hydrodynamic equations in a system with broken time reversal symmetry are equipped with a non-dissipative viscosity coefficient, termed odd viscosity. Such odd viscosity was experimentally measured in a suspension of particles driven by an external field \cite{Soni2019}. Active fluids composed of self-propelling particles\cite{PhysRevE.99.042605,PhysRevE.101.022603} with broken time reversal symmetry have thus attracted great interest. The constant injection of external energy leads to non-equilibrium states of active particles, where diverse steady-state collective motions sharply contrast with those commonly observed in equilibrium systems\cite{Elgeti2015,Zoettl2016,Sabrina2018,Shaebani2020,Qi2022,Lei2019b,Zhang2022}. Most of the research about active colloidal particles has focused on the active colloidal particles themselves and limited attention has been put on the medium, which mediates the hydrodynamic interactions between active colloidal particles. Nearly all the collective behavior of active colloidal particles is studied in a liquid environment. This makes hydrodynamic interactions  essential\cite{Marchetti2013,Zoettl2014} for understanding phenomena observed in active systems. More importantly, due to the constant input of external energy in active systems, the mechanical motions of embedded active units also constantly disturb the surrounding liquid. These induced active flows are absent in usual passive fluids.

In this work, we envision a collection of particles suspended in an active fluid, which possesses both even and odd viscosity. In an analogy to the development of traditional Stokesian Dynamics, we systematically derive an extended Stokesian Dynamics formalism to explore possible physical effects of odd viscosity\cite{Khain2022,Samanta2022,Han2021a,PhysRevE.104.064613}. We first briefly review the concept of odd viscosity in section \ref{sec-odd-viscosity}. Then, the Stokes' equations governing the conventional micro-hydrodynamics are generalized to include the additional effects caused by odd viscosity in section \ref{sec-stokes-odd} and the Green's function of the generalized Stokes' equations is also derived in section \ref{sec-ossen-odd}. By considering up to the first order effect of odd viscosity in section \ref{sec-sd-main}, both near-field and far-field mobility tensors due to odd viscosity are calculated in section \ref{sec-mobility-tensors-near} and section \ref{sec-mobility-tensors-far}, respectively. In section \ref{sec-mobility-tensors-full}, those results are incorporated into the existing Stokesian Dynamics framework for simulating a population of particles suspended in an active fluid medium with both even and odd viscosity. The whole work concludes with a brief summary and discussions about possible future work.

\section{Stokes' equations with odd viscosity}

\subsection{Odd viscosity}\label{sec-odd-viscosity}
Readers already familiar with odd viscosity can skip this section, where the concept of odd viscosity is briefly reviewed\cite{avron1998odd,PhysRevE.103.042610,PhysRevLett.127.048001,Markovich2019,Banerjee2017}.  Without loss of generality, the viscous stresses of any fluids in response to external disturbances can be written as
\begin{equation}
    \sigma_{ij}=-p\delta_{ij}+\mu_{ijkl}\nabla_l u_k
\end{equation}
where $\sigma_{ij}$ is the stress tensor, $p$ is the isotropic hydrostatic pressure and $\nabla_l u_k$ is the velocity gradient tensor. Here, $\mu_{ijkl}$ is the fourth order viscosity tensor, which describes resulting viscous stresses in response to velocity gradients. The above relation is usually termed as the linear constitutive relation of a fluid. Analogous to the case of a homogeneous isotropic linear elastic solid \cite{Krempl2009}, the spatial symmetry properties of a homogeneous isotropic fluid simplifies the general viscosity tensor to 
\begin{equation}
    \mu_{ijkl}^e=\mu_b \delta_{ij}\delta_{kl} + \mu_s (\delta_{ik}\delta_{jl}+\delta_{jk}\delta_{il})
    \label{eqn-viscosity-even}
\end{equation}
where $\mu_b$ and $\mu_s$ are the bulk and shear viscosity, respectively. Such viscosity tensor $\mu_{ijkl}^e$ only depends on two independent material properties, which play similar roles of bulk and shear modulus in linear elastic solids.

Furthermore, symmetry properties of the system enforce three important symmetry relations, which the viscosity tensor $\mu_{ijkl}^e$ must fulfill. Firstly, the rotational invariance of system requires that $\mu^e_{ijkl}=\mu^e_{ijlk}$. Secondly, the conservation of local angular momentum (no external torques) requires that the stress tensor must be symmetric, that is,  $\sigma_{ij}=\sigma_{ji}$, which enforces that $\mu^e_{ijkl}=\mu^e_{jikl}$. Thirdly, according to the Onsager reciprocal relations\cite{PhysRev.37.405,Landau2013}, the time reversal symmetry requires that $\mu^e_{ijkl}=\mu^e_{klij}$. It is easy to verify that the above viscosity expression $\mu_{ijkl}^e$ fulfills all three symmetry relations. The superscript $e$ indicates that it is even under the time-reversal transformation ($ij\leftrightarrow kl$) and the corresponding viscosity $\mu_{ijkl}^e$ is therefore termed as \emph{even viscosity}.

As an extension of the above symmetry considerations, Avron et al.\cite{avron1998odd} considered a system under an external magnetic field, which possesses a broken time-reversal symmetry. In such a scenario, the Onsager reciprocal relations can be further extended for systems with a broken time reversal symmetry\cite{Groot1984}, which yields
\begin{equation}
    \mu_{ijkl}^o = - \mu_{klij}^o
\end{equation}
where the superscript $o$ indicates that the viscosity is odd under the time-reversal transformation and $\mu_{ijkl}^o$ is thus analogously termed as \emph{odd viscosity}.

The most general expression of odd viscosity in a 3D system can be quite complex and may contain numerous components depending on the symmetry of system, which is systematically explored by Vitelli et al.\cite{Khain2022}. Here, we restrict our attention to a more specific case by only considering the possible rotational degrees of freedom of microscopic fluid particles. Starting from a microscopic Hamiltonian, Lubensky et al. \cite{PhysRevLett.127.048001} derived an expression of odd viscosity in a 3D system as,
\begin{equation}
    \mu_{ijkl}^o=\ell_n (\varepsilon_{jln}\delta_{ik}+\varepsilon_{iln}\delta_{jk}+\varepsilon_{ikn}\delta_{jl}+\varepsilon_{jkn}\delta_{il})
    \label{eqn-viscosity-odd}
\end{equation}
where $\varepsilon_{ijk}$ is the three-dimensional (3D) Levi-Civita symbol, $\delta_{ij}$ is the Kronecker delta, and $\ell_i$ is the local angular momentum density of constituting particles. Please be aware that the expression above derived in the work \cite{PhysRevLett.127.048001} matches with the general expression derived in the work\cite{Khain2022} by taking $\eta_1^0=-2\eta_2^0=\mu_0$ in their notations. And similar to the case of even viscosity, the odd viscosity should fulfill three symmetry relations $\mu_{ijkl}^o=\mu_{ijlk}^o$, $\mu_{ijkl}^o=\mu_{jikl}^o$ and $\mu_{ijkl}^o=-\mu_{klij}^o$. It can be easily verified that all three symmetric relations are fulfilled by Eqn. \ref{eqn-viscosity-odd}. For a quasi-2D system where the local angular momentum density is constrained to be uni-axial such as $\gv \ell=\mu_o \uv z$, the above general expression of odd viscosity in 3D (Eqn. \ref{eqn-viscosity-odd}) can be further simplified to
\begin{equation}
    \mu_{ijkl}^o=\mu_o (\epsilon_{jl}\delta_{ik}+\epsilon_{il}\delta_{jk}+\epsilon_{ik}\delta_{jl}+\epsilon_{jk}\delta_{il})
    \label{eqn-viscosity-odd-2d}
\end{equation}
where $\varepsilon_{ij}$ is the 2D Levi-Civita symbol defined as $\varepsilon_{ij}=\varepsilon_{ijz}$ and the index of $\epsilon_{ij}$ is extended to run over x, y and z. In such a simplified case, a single scalar material parameter --- odd viscosity $\mu_o$ --- is sufficient to determine the whole odd viscosity tensor $\mu_{ijkl}^o$.

Finally, for a fluid involving both even and odd interactions under the time-reversal transformation, the general linear constitutive relation of such a fluid can be expressed as
\begin{equation}
    \sigma_{ij}=-p\delta_{ij}+\left(\mu_{ijkl}^e+\mu_{ijkl}^o\right)\nabla_l u_k
    \label{eqn-material-model}
\end{equation}
where the explicit expressions of even and odd viscosity tensors are given by Eqn. \ref{eqn-viscosity-even} and \ref{eqn-viscosity-odd}, respectively.

\subsection{Odd Stokes' equations}\label{sec-stokes-odd}
With the constitutive relation of a fluid including both even and odd viscosity (Eqn. \ref{eqn-material-model}), the corresponding Stokes' equations are ready to be derived via the balance of linear momentum. For simplicity, we only consider quasi-2D systems here and the overall viscous stress tensor is thus given by
\begin{equation}
\begin{aligned}
    \sigma_{ij}=-p\delta_{ij}&+\mu_b \nabla_k u_k \delta_{ij} + \mu_s\left(\nabla_j u_i + \nabla_i u_j\right)\\
    &+\mu_o \left(\varepsilon_{jl}\nabla_l u_i+\varepsilon_{il} \nabla_l u_j +\nabla_i \varepsilon_{jl}u_l+\nabla_j \varepsilon_{il}u_l\right)
\end{aligned}
\end{equation}
For an incompressible fluid, $\nabla_j u_j =0$, the balance of internal forces $\left(\div{\gv \sigma}\right)$ and external forces ($\v f$) yields the Stokes' equations with odd viscosity, 
\begin{equation}
    -\nabla_i \left(p - \mu_o \epsilon_{jl}\nabla_j u_l\right) +\mu_s \nabla^2 u_i + \mu_o \epsilon_{il} \nabla^2 u_l=f_i
\end{equation}
The above equations can also be rewritten in an equivalent matrix form as
\begin{equation}
    \begin{pmatrix}
    \mu_s&\mu_o&0\\
    -\mu_o&\mu_s&0\\
    0&0&\mu_s
    \end{pmatrix}
    \nabla^2 \v u = \grad \tilde{p} + \v f
    \label{eqn-Stokes}
\end{equation}
where the modified pressure $\tilde{p}$ is defined as $\tilde{p}=p - \mu_o \epsilon_{jl}\nabla_j u_l$ and $\v f$ denotes possible external volumetric forces such as gravity. The above equation reduces to the conventional Stokes' equation when $\mu_o=0$ and the odd viscosity $\mu_o$ emerges only at off-diagonal elements, which introduces additional couplings between in-plane components (x and y components here, with angular momentum density $\gv \ell$ assumed to be along z axis).

\subsection{Odd Oseen tensor}\label{sec-ossen-odd}
As the Stokes' equations with odd viscosity (Eqn. \ref{eqn-Stokes}) are still linear equations, here we derive its corresponding Green's function. Similarly to the Oseen tensor\cite{SangtaeKim2005,Graham2018} derived for the conventional Stokes' equations with only even viscosity, we start by considering a point force $-\v F$ applied at position $\v r$\cite{Morse1953}, that is, 
\begin{equation}
    \begin{pmatrix}
    \mu_s&\mu_o&0\\
    -\mu_o&\mu_s&0\\
    0&0&\mu_s
    \end{pmatrix}
    \nabla^2 \v u = \grad p - \v F \delta(\v r)
    \label{eqn-Stokes-pf}
\end{equation}
Here, the modified pressure $\tilde{p}$ (Eqn. \ref{eqn-Stokes}) is denoted as $p$ for brevity. Due to the linearity of the odd Stokes' equations, both the pressure field $p$ and flow field $\v u$ should be proportional to the applied point force $\v F$. Since $p(\v r)$ is a scalar field and $\v u(\v r)$ is a vector field, such linear relations can be expressed as
\begin{equation}
    p(\v r) = P_j(\v r) F_j, \quad
    u_i(\v r) = G_{ij}(\v r) F_j
\label{eqn-linearity}
\end{equation}
where $P_i(\v r)$ and $G_{ij}(\v r)$ are the Green's function of the pressure field and flow fields, respectively. Applying the Fourier transformation to Eqn. \ref{eqn-Stokes-pf}, its corresponding expressions in k-space can be computed as
\begin{equation}
    -\mu_{ij}k^2 \widehat{u}_j(\v k) = i k_i \widehat{p}(\v k) - F_i
    \label{eqn-Stokes-pf-kspace}
\end{equation}
where $\v k$ is the wave vector, $\widehat{p}(\v k)$ and $\widehat{\v u}(\v k)$ are the respective pressure field and flow fields in the k-space. For convenience, a second order auxiliary viscosity matrix $\mu_{ij}$ is introduced, 
\begin{equation}
    \mu_{ij}=
    \begin{pmatrix}
    \mu_s&\mu_o&0\\
    -\mu_o&\mu_s&0\\
    0&0&\mu_s
    \end{pmatrix}
    =\mu_s\delta_{ij} + \mu_o \varepsilon_{ij}
    \label{eqn-viscosity-rank2}
\end{equation}
Utilizing the linearity of odd Stokes' equations (Eqn. \ref{eqn-linearity}) and the incompressibility condition of fluids ($\div{\v u}=0$), the Green's functions of both pressure and flow field can be solved analytically in k-space as (see Appx. \ref{appx-Green-func-k-space}):
\begin{equation}
    \widehat{P}_j(\v k) = - \frac{ik_j}{k^2} + i \mu_0 \varepsilon_{ik}k_i \widehat{G}_{kj}(\v k), \quad \widehat{G}_{ij}(\v k) = \widehat{G}_{(ij)}(\v k) + \widehat{G}_{[ij]}(\v k)
    \label{eqn-Grenn-func-k}
\end{equation}
where $\widehat{P}_j(\v k)$ and $\widehat{G}_{ij}(\v k)$ are the corresponding Green's functions of pressure and flow fields in the k-space. The k-space Green's function of flow fields $\widehat{G}_{ij}(\v k)$ can be split as the sum of its symmetric and anti-symmetric parts:
\begin{equation}
    \widehat{G}_{(ij)}(\v k)=\frac{1}{\mu_s}\frac{1}{\Sigma^2(\v k)}\left(-\frac{k_ik_j}{k^4}+\frac{\delta_{ij}}{k^2}\right), \quad \widehat{G}_{[ij]}(\v k)=-\frac{1}{\mu_s} \frac{\chi}{\Sigma^2(\v k)}\varepsilon_{ijk}k_k \frac{k_z}{k^4}
    \label{eqn-Green-func-k-decomposed}
\end{equation}
where two auxiliary variables introduced are defined as:
\begin{equation}
    \Sigma^2(\v k) = 1+\chi^2\frac{k_z^2}{k^2}, \quad \chi = \frac{\mu_o}{\mu_s}
\end{equation}
Particularly, the dimensionless number $\chi$ is the ratio between odd viscosity $\mu_o$ and even viscosity $\mu_s$, which naturally characterizes the significance of odd viscosity related effects relative to even viscosity.

The corresponding real-space expressions of Green's functions can be obtained via the inverse Fourier transformation of its k-space expressions (Eqn. \ref{eqn-Grenn-func-k}). However, the additional k-dependence introduced by $\Sigma^2(\v k)$ for non-vanishing odd viscosity makes the exact analytical calculation difficult. To avoid obscuring essential physics by tedious algebra, we only focus on the fluids with a weak odd viscosity\cite{Soni2019}, that is, $\mu_o \ll \mu_s$. Then, the viscosity ratio $\chi \ll 1 $ serves as a natural expansion parameter for obtaining the leading effects of odd viscosity (see Appx. \ref{appx-Green-func-real-space}). The real-space expressions of the Green's functions can be similarly decomposed into symmetric and anti-symmetric parts, $G_{ij}(\v r) = G_{(ij)}(\v r) + G_{[ij]}(\v r)$, which can be systematically expanded in powers of $\chi$ as
\begin{equation}
    G_{(ij)}(\v r) = G_{(ij)}^{(0)}(\v r) + G_{(ij)}^{(2)}(\v r) + \mathcal{O}(\chi^4), \quad
    G_{[ij]}(\v r) = G_{[ij]}^{(1)}(\v r) + G_{[ij]}^{(3)}(\v r) + \mathcal{O}(\chi^5)
    \label{eqn-Green-func-real-main}
\end{equation}
where $G_{(ij)}(\v r)$ and $G_{[ij]}(\v r)$ are the symmetric and anti-symmetric parts of the real-space Green's functions $G_{ij}(\v r)$, respectively. The number inside the superscript parenthesis indicates powers of the viscosity ratio $\chi$. The explicit expressions of the symmetric Green's functions are given by Eqn. \ref{eqn-Green-func-sym-0} and \ref{eqn-Green-func-sym-2} and the explicit expressions of the anti-symmetric Green's functions are given by Eqn. \ref{eqn-Green-func-antisym-1} and \ref{eqn-Green-func-antisym-3}. Please refer to Appx. \ref{appx-Green-func-real-space} for the detailed derivations.

\section{Stokesian Dynamics with odd viscosity}\label{sec-sd-main}
With the Green's function (Eqn. \ref{eqn-Green-func-real-main}) of the odd Stokes' equations (Eqn. \ref{eqn-Stokes}) derived in previous sections, the flow fields induced by a point force $\v F$ at origin can be simply expressed as
\begin{equation}
    u^{PF}_i(\v r) = G_{ij}(\v r) F_j=\left[G_{(ij)}(\v r) + G_{[ij]}(\v r)\right] F_j
\end{equation}
For simplicity, only the first order effect of odd viscosity will be explored in this work but the formalism developed in this section in principle can be extended to arbitrarily higher orders. If we retain up to the first order terms of viscosity ratio $\chi$ (Eqn. \ref{eqn-Green-func-sym-0} and \ref{eqn-Green-func-antisym-1}), the first order approximation of the exact Green's function is
\begin{equation}
\begin{aligned}
    G_{ij}(\v r) = G_{(ij)}^{(0)}(\v r) + G_{[ij]}^{(1)}(\v r) + \mathcal{O}(\chi^2)
    \approx \frac{1}{8\pi\mu_s}\left[J_{ij}(\v r)-\chi J_{ij}^o(\v r)\right]
\end{aligned}
\end{equation}
where $J_{ij}(\v r)$ is the conventional \emph{Ossen tensor} (Eqn. \ref{eqn-Oseen}) derived from Stokes' equations with only even viscosity\cite{SangtaeKim2005}. $J_{ij}^o(\v r)$ is the additional response due to the existence of odd viscosity, which in analogy to the conventional terminology used for even viscosity, we referred to it as \emph{odd Oseen tensor} (Eqn. \ref{eqn-Oseen-odd}). Hereafter, the shear viscosity $\mu_s$ will be denoted as $\mu$ for short. Comparing with the conventional passive fluids, the above Green's function has one additional term due to the first order effect of the odd viscosity $-\chi J^o_{ij}(\v r)$. By the linearity of the odd Stokes' equations (Eqn. \ref{eqn-Stokes}), the total flow fields induced by a collection of $N$ finite-sized rigid particles can be written as
\begin{equation}
    u_i(\v x) - u_i^\infty(\v x) = \frac{1}{8\pi\mu}\sum_{\alpha=1}^N\oint_{S^\alpha} \left[J_{ij}(\v r)-\chi J_{ij}^o(\v r)\right] f_j(\v y) \mathrm{d} S(\v y)
    \label{eqn-integral-req}
\end{equation}
where $u_i^\infty$ ($i=x,y,z$) is the background linear velocity field, $S^\alpha$ is the surface of $\alpha$-th particle ($\alpha=1,\cdots,N$), $\v y$ denotes any locations at the particle surface, and $f_j(\v y)$ is the force density distributed over the particle surface.

Based on the above integral representation of odd Stokes' equations, we will extend the conventional Stokesian Dynamics formalism to include the effects due to odd viscosity. Since the Stokesian Dynamics with only even viscosity has been extensively studied\cite{Brady1988}, we will only focus on the additional flows contributed by odd viscosity, which is denoted as $u_i^o(\v x)$ and
\begin{equation}
    u_i^o(\v x) - u_i^\infty(\v x)=-\frac{\chi}{8\pi\mu}\sum_{\alpha=1}^N\oint_{S^\alpha}  J_{ij}^o(\v r) f_j(\v y) \mathrm{d} S(\v y)
    \label{eqn-integral-req-odd}
\end{equation}

\subsection{Multipole expansion}
Commonly speaking, the detailed information of force density distribution on each particle's surface is difficult to obtain and usually only moments of force density distribution, such as forces, torques and so on, are experimentally accessible. Therefore, induced flows in response to each force density moment are of more interest in practice. If we only consider up to the first order moments of force density distribution, the integral form expression of the induced odd flows (Eqn. \ref{eqn-integral-req-odd}) can be systematically expanded via the multipole expansion
\begin{equation}
    u_i^o(\v x) - u_i^\infty(\v x)\approx-\frac{\chi}{8\pi\mu}\sum_{\alpha=1}^N \left[J_{ij}^o(\v x - \v x^\alpha)F_j^\alpha + R^o_{ij}(\v x - \v x^\alpha) T_j^\alpha - K^o_{ijk}(\v x - \v x^\alpha)S_{jk}^\alpha\right]
    \label{eqn-multipole-odd}
\end{equation}
where $\v F^\alpha$, $\v T^\alpha$ and $\ten{S}^\alpha$ are the total forces, torques and stresslets acting on the $\alpha$-th particle, respectively, which are the zeroth and first order moments of force density distribution\cite{Guazzelli2011}; $J^o_{ij}(\v r)$, $R^o_{ij}(\v r)$ and $K^o_{ijk}(\v r)$ are the respective odd viscosity relevant propagators of forces, torques and stresslets, which are defined as
\begin{subequations}
    \begin{align}
    J_{ij}^o(\v r)&=\frac{\varepsilon_{ij3}}{r} - \frac{\varepsilon_{ijk}r_kr_3}{r^3}\\
    R_{ij}^o(\v r)&=\frac{1}{2}\varepsilon_{ikl}\nabla_{k} J_{jl}^o\\
    K_{ijk}^o(\v r)&=\frac{1}{2}\left[\nabla_k J_{ij}^o(\v r) + \nabla_j J_{ik}^o(\v r)\right]
    \end{align}
    \label{eqn-propagators-odd}
\end{subequations}
The above multipole expansion (Eqn. \ref{eqn-multipole-odd}) is truncated at the first order although such expansion can proceed to higher order terms for a better accuracy. Particularly, for rigid spherical particles, such expansion terminates very quickly and even an exact solution can be obtained for isolated particles just like the case with only even viscosity \cite{SangtaeKim2005}. Without loss of generality, the induced odd flow fields can be expressed as
\begin{equation}
    u_i^{o} (\v x) - u_i^\infty(\v x) \approx -\frac{\chi}{8\pi\mu} \sum_{\alpha=1}^N \left\{\mathcal{F}\left[J^o_{ij}(\v x-\v x^\alpha)\right] F_j^\alpha + \mathcal{T}\left[J^o_{ij}(\v x-\v x^\alpha)\right] T_j^\alpha + \mathcal{S}\left[J^o_{ij}(\v x-\v x^\alpha)\right] S_{jk}^\alpha\right\}
\end{equation}
where $\mathcal{F}\left[J^o_{ij}(\v x-\v y)\right]$, $\mathcal{T}\left[J^o_{ij}(\v x-\v y)\right]$ and $\mathcal{S}\left[J^o_{ij}(\v x-\v y)\right]$ denote linear functions which describe the odd viscosity related responses to external forces, torques and stresslets, respectively. If the particles are considered ideal point particles, the explicit functional forms of those response functions are just the propagators of each force moment (Eqn. \ref{eqn-propagators-odd}). However, due to the fact that particles are of finite size, there are additional non-vanishing higher order force moments that need to be considered.

Next, we determine the exact functional forms of the response functions with higher order terms. We first extend the Lorentz reciprocal theorem and the Fax\'en laws to systems with odd viscosity. Then, the additional mobility tensors due to odd viscosity are systematically derived in both near and far fields for the development of Stokesian Dynamics with odd viscosity.

\subsection{Near-field mobility tensors}\label{sec-mobility-tensors-near}
\subsubsection{Ambient flows in integral form}
Here, we restrict our consideration to spherical rigid particles. Their spherical symmetry makes it reasonable to assume the force moments are uniformly distributed over the particle surface. Then, the integral form of ambient odd flows (odd flows induced by other particles excluding $\alpha$-th particle itself) can be written as
\begin{equation}
\begin{aligned}
    u_i^{o\prime}(\v x) - u_i^\infty(\v x)&\approx-\frac{\chi}{8\pi\mu}\sum_{\beta=1,\beta\neq\alpha}^N\oint_{S^\beta} J^o_{ij}(\v x - \v y)\left[\frac{F_j^\beta}{4\pi a^2}+\frac{3}{8\pi a^3}\epsilon_{jkl}T_k^\beta n_l\mathrm{d}S+\frac{3}{4\pi a^3}S_{jk}^\beta n_k\right]\mathrm{d}S\\
\end{aligned}
\label{eqn-ambient-flow-integral-odd}
\end{equation}
where $u_i^{o\prime}$ denotes the ambient odd flows. Such an approximation is closely related to the idea of the Rotne-Prager-Yamakawa (RPY) approximation\cite{Rotne1969,Yamakawa1970,Yamakawa1971} commonly used in conventional Stokesian Dynamics.
Then, the integral forms of response functions $\mathcal{F}\left[J^o_{ij}(\v x-\v y)\right]$, $\mathcal{T}\left[J^o_{ij}(\v x-\v y)\right]$ and $\mathcal{S}\left[J^o_{ij}(\v x-\v y)\right]$ can be defined as
\begin{subequations}
\begin{align}
    \mathcal{F}\left[J^o_{ij}(\v x-\v y);S\right]&=\frac{1}{4\pi a^2}\oint_{S} \widetilde{J}^o_{ij}(\v x - \v y)\mathrm{d}S(\v y)\\
    \mathcal{T}\left[J^o_{ij}(\v x-\v y);S\right]&=\frac{3}{8\pi a^3}\oint_{S} \widetilde{R}^o_{ij}(\v x - \v y) \mathrm{d}S(\v y)\\
    \mathcal{S}\left[J^o_{ij}(\v x-\v y);S\right]&=\frac{3}{4\pi a^3}\oint_{S} \widetilde{K}^o_{ijk}(\v x - \v y)\mathrm{d}S(\v y)
\end{align}
\label{eqn-response-integral}
\end{subequations}
where integrals are evaluated at the particle surface $S$ and integrands inside surface integrals of each force moments are defined as
\begin{subequations}
\begin{align}
    \widetilde{J}_{ij}^o(\v x-\v y)&=J^o_{ij}(\v x - \v y)\\
    \widetilde{R}_{ij}^o(\v x-\v y)&=J^o_{ik}(\v x - \v y)\varepsilon_{kjl}n_l\\
    \widetilde{K}_{ijk}^o(\v x-\v y)&=\frac{1}{2}\left[J^o_{ij}(\v x - \v y)n_k + J^o_{ik}(\v x - \v y)n_j\right]
\end{align}
\end{subequations}
For spherical rigid particles, the single spherical surface integrals of response functions can be evaluated analytically, which yield
\begin{subequations}
\begin{equation}
    M^o_{Fij}(\v x)= \frac{1}{8\pi\mu}\mathcal{F}\left[J^o_{ij}(\v x-\v y);S^\beta \right]=
    \begin{cases}
    \left(1+\frac{a^2}{6}\nabla^2\right) \frac{J^o_{ij}(\v x - \v x^\beta)}{8\pi\mu} &\abs{\v x - \v x^\beta}>a\\
    \frac{1}{12\pi\mu a} \varepsilon_{ij3}&\abs{\v x - \v x^\beta} \le a
    \end{cases}
\end{equation}

\begin{equation}
    M^o_{Tij}(\v x) = \frac{1}{8\pi\mu}\mathcal{T}\left[J^o_{ij}(\v x-\v y);S^\beta\right]=
    \begin{cases}
    -\left(1+\frac{a^2}{10}\nabla^2\right)\frac{R_{ij}^o(\v x - \v x^\beta)}{8\pi\mu}&\abs{\v x - \v x^\beta}>a\\
    \frac{1}{8\pi\mu a^3}\left[-\frac{3}{5}\left(x_l-x^\beta_l\right)\delta_{lij3}\right.&\\
    \left.-\frac{1}{2}\delta_{ij}\left(z-z^\beta\right) + \frac{1}{2} \delta_{i3}(x_j - x^\beta_j)\right] & \abs{\v x - \v x^\beta} \le a
    \end{cases}
\end{equation}

\begin{equation}
    M^o_{Sijk}(\v x)=\frac{1}{8\pi\mu}\mathcal{S}\left[J^o_{ij}(\v x-\v y);S^\beta\right]=
    \begin{cases}
    -\left(1+\frac{a^2}{10}\nabla^2\right)\frac{K^o_{ijk}(\v x - \v x^\beta)}{8\pi\mu}&\abs{\v x - \v x^\beta} > a\\
    \frac{1}{40\pi\mu a^3}\left\{\left[\varepsilon_{ij3}\left(x_k - x_k^\beta\right)+\varepsilon_{ik3}\left(x_j - x_j^\beta\right)\right] \right.& \\
    \left.+ \left(\varepsilon_{ijm}\delta_{k3}+\varepsilon_{ikm}\delta_{j3}\right)\left(x_m - x_m^\beta\right)\right\}&\abs{\v x - \v x^\beta} \le a
    \end{cases}
\end{equation}
\label{eqn-ambient-flow-integral-overlap-odd}
\end{subequations}

\subsubsection{The Fax\'en laws in integral form}
In the derivation of a traditional Stokesian Dynamics formalism, the Fax\'en laws\cite{SangtaeKim2005,Guazzelli2011} provide a bridge for connecting the kinetic motions $(\v U^\alpha, \v \Omega^\alpha, \ten{E}^\alpha)$ with corresponding mechanical quantities $(\v F^\beta, \v T^\beta, \ten{S}^\beta)$ of each particles. However, the validity of the Fax\'en laws with even viscosity\cite{SangtaeKim2005} relies on the Lorentz reciprocal theorem and symmetry properties of the Oseen tensor. Due to the existence of odd viscosity, we can derive the generalized Lorentz reciprocal theorem (Appx. \ref{appx-lrt}) based on the symmetry properties of even and odd viscosity. With the generalized Lorentz reciprocal theorem and symmetry properties of the odd Oseen tensor, a generalized version of the Fax\'en laws including both even and odd viscosity are derived (Appx. \ref{appx-faxen-laws}). Comparing with the conventional Fax\'en laws, there is an additional minus sign for the terms contributed by odd viscosity in the generalized Fax\'en laws.

With the integral forms of response functions (Eqn. \ref{eqn-response-integral}) and the generalized Fax\'en laws for odd viscosity (Eqn. \ref{eqn-Faxen-laws-odd}), the integral form of Fax\'en laws with odd viscosity can be written explicitly as 
\begin{equation}
\left\{
\begin{aligned}
U_i^\alpha - U_i^\infty(\v x^\alpha) &=- \mathcal{F}\left[u_i^{o\prime}(\v x);S^\alpha\right]=-\frac{1}{4\pi a^2}\oint_{S^\alpha}u_i^{o\prime}(\v x)\mathrm{d}S^\alpha(\v x)\\
\Omega_i^\alpha - \Omega_i^\infty(\v x^\alpha)&=-\mathcal{T}\left[u_i^{o\prime}(\v x);S^\alpha\right]=-\frac{3}{8\pi a^3}\oint_{S^\alpha}\varepsilon_{ijk}n_ju_k^{o\prime}(\v x)\mathrm{d}S^\alpha(\v x)\\
E_{ij}^\alpha-E_{ij}^\infty(\v x^\alpha)&=-\mathcal{S}[u_i^{o\prime}(\v x);S^\alpha]=-\frac{3}{4\pi a^3}\oint_{S^\alpha}\frac{1}{2}\left[u^{o\prime}_i(\v x)n_j+u^{o\prime}_j(\v x)n_i\right]\mathrm{d}S^\alpha(\v x)
\end{aligned}
\right.
\label{eqn-Faxen-integral-odd}
\end{equation}
where $\mathcal{F}\left[u_i^{o\prime}(\v x);S^\alpha\right]$, $\mathcal{T}\left[u_i^{o\prime}(\v x);S^\alpha\right]$ and $\mathcal{S}\left[u_i^{o\prime}(\v x);S^\alpha\right]$ are the same linear functions defined by Eqn. \ref{eqn-response-integral} but applied to the ambient odd flow $\v u^{o\prime}(\v x)$ given by Eqn. \ref{eqn-ambient-flow-integral-odd} and evaluated at the surface of $\alpha$-th particle $S^\alpha$ instead.

\subsubsection{Near-field mobility tensors}
Then, the integral form pair mobility tensors ($\alpha\neq\beta$) can be obtained by plugging the expression of ambient flow (Eqn. \ref{eqn-ambient-flow-integral-odd}) into the Fax\'en laws in integral form (Eqn. \ref{eqn-Faxen-integral-odd}), which yields 
\begin{subequations}
\begin{align}
    \ten{M}_{UF}^{\alpha\beta}=\left(M\indices{^o_{UF}^{\alpha\beta}_{ij}}\right)&=\frac{\chi}{8\pi\mu}\frac{1}{\left(4\pi a^2\right)^2}\oint_{S^\alpha}\oint_{S^\beta} \widetilde{J}^o_{ij}(\v x - \v y) \mathrm{d}S^\alpha\mathrm{d}S^\beta\\
    \ten{M}_{UT}^{\alpha\beta}=\left(M\indices{^o_{UT}^{\alpha\beta}_{ij}}\right)&=\frac{\chi}{8\pi\mu}\frac{1}{4\pi a^2}\frac{3}{8\pi a^3}\oint_{S^\alpha}\oint_{S^\beta} \widetilde{R}^o_{ij}(\v x - \v y) \mathrm{d}S^\alpha\mathrm{d}S^\beta\\
    \ten{M}_{US}^{\alpha\beta}=\left(M\indices{^o_{US}^{\alpha\beta}_{ijk}}\right)&=\frac{\chi}{8\pi\mu}\frac{1}{4\pi a^2}\frac{3}{4\pi a^3}\oint_{S^\alpha}\oint_{S^\beta} \widetilde{K}^o_{ijk}(\v x - \v y) \mathrm{d}S^\alpha\mathrm{d}S^\beta\\
    \ten{M}_{\Omega T}^{\alpha\beta}=\left(M\indices{^o_{\Omega T}^{\alpha\beta}_{ij}}\right)&=\frac{\chi}{8\pi\mu}\left(\frac{3}{8\pi a^3}\right)^2\oint_{S^\alpha}\oint_{S^\beta}\varepsilon_{ikl}n_k\widetilde{R}^o_{lj}(\v x - \v y)\mathrm{d}S^\alpha\mathrm{d}S^\beta\\
    \ten{M}_{\Omega S}^{\alpha\beta}=\left(M\indices{^o_{\Omega S}^{\alpha\beta}_{ijk}}\right)&=\frac{\chi}{8\pi\mu}\frac{3}{8\pi a^3}\frac{3}{4\pi a^3}\oint_{S^\alpha}\oint_{S^\beta}\varepsilon_{ilm}n_l\widetilde{K}^o_{mjk}(\v x - \v y)\mathrm{d}S^\alpha\mathrm{d}S^\beta\\
    \ten{M}_{E S}^{\alpha\beta}=\left(M\indices{^o_{ES}^{\alpha\beta}_{ijkl}}\right)&=\frac{\chi}{8\pi\mu}\left(\frac{3}{4\pi a^3}\right)^2
    \oint_{S^\alpha}\oint_{S^\beta}\frac{1}{2}\left[\widetilde{K}^o_{ikl}(\v x - \v y)n_j+\widetilde{K}^o_{jkl}(\v x - \v y)n_i\right]\mathrm{d}S^\alpha\mathrm{d}S^\beta
\end{align}
\label{eqn-mobility-tensors-integral-odd}
\end{subequations}
The evaluation of the double surface integrals are tedious and its final explicit expressions of near-field \emph{odd} mobility tensors are documented in Appx. \ref{appx-sd-mobility-near-pair-odd}.

\subsubsection{Self-mobility tensors}
Traditionally, the self-mobility tensors are derived from the analytical solutions of each mode of motion for an isolated particle. However, starting from only Green's functions, the RPY approach provides a way to derive the self-mobility tensors without solving each boundary-value problem as the near-field mobility tensors (Appx. \ref{appx-sd-mobility-near-pair-odd}) should continuously reduce to the self-mobility tensors as the separation distance goes to zero. Below, the self-mobility tensors (Appx. \ref{appx-sd-mobility-self-odd}) are calculated by taking the limits of zero separation distance of the near-field pair mobility tensors
\begin{subequations}
\begin{align}
    M\indices{^o_{UF}^{\alpha\alpha}_{ij}}&=\lim_{r\to0}M\indices{^o_{UF}^{\alpha\beta}_{ij}}=\frac{\chi}{12\pi\mu a}\varepsilon_{ij3}\\
    M\indices{^o_{UT}^{\alpha\alpha}_{ij}}&=\lim_{r\to0}M\indices{^o_{UT}^{\alpha\beta}_{ij}}=0\\
    M\indices{^o_{US}^{\alpha\alpha}_{ijk}}&=\lim_{r\to0}M\indices{^o_{US}^{\alpha\beta}_{ijk}}=0\\
    M\indices{^o_{\Omega T}^{\alpha\alpha}_{ij}}&=\lim_{r\to0}M\indices{^o_{\Omega T}^{\alpha\beta}_{ij}}=\frac{\chi}{16\pi\mu a^3}\varepsilon_{ij3}\\
    M\indices{^o_{\Omega S}^{\alpha\alpha}_{ijk}}&=\lim_{r\to0}M\indices{^o_{\Omega S}^{\alpha\beta}_{ijk}}=\frac{3\chi}{40\pi\mu a^3}\delta_{3ijk}\\
    M\indices{^o_{ES}^{\alpha\alpha}_{ijkl}}&=\lim_{r\to0}M\indices{^o_{ES}^{\alpha\beta}_{ijkl}}=\frac{\chi}{80\pi\mu a^3}\left[\left(\varepsilon_{ik3}\delta_{lj}+\varepsilon_{il3}\delta_{kj}\right) + \left(\varepsilon_{jk3}\delta_{li}+\varepsilon_{jl3}\delta_{ki}\right)\right]
\end{align}
\label{eqn-mobility-tensors-self-odd}
\end{subequations}

\subsection{Far-field mobility tensors}\label{sec-mobility-tensors-far}

\subsubsection{Ambient flows in differential form}
With the integral form of ambient flow (Eqn. \ref{eqn-ambient-flow-integral-odd}) and the far-field results of the single surface integrals of propagators (Eqn. \ref{eqn-ambient-flow-integral-overlap-odd}, $\abs{\v x - \v x^\beta}>a$), we can show that the far-field expression of the near-field ambient flow (Eqn. \ref{eqn-ambient-flow-integral-odd}) is
\begin{equation}
\begin{aligned}
    u^{o\prime}_i(\v x) - u_i^\infty(\v x)\approx
    &-\frac{\chi}{8\pi\mu}\sum_{\beta=1,\beta\neq\alpha}^N \left[\left(1+\frac{1}{6}a^2\nabla^2\right)J_{ij}^o(\v x - \v x^\beta)F_j^\beta -\left(1+\frac{1}{10}a^2\nabla^2\right) R^o_{ij}(\v x - \v x^\beta) T_j^\beta \right.\\
    &\left.- \left(1+\frac{1}{10}a^2\nabla^2\right)K^o_{ijk}(\v x - \v x^\beta)S_{jk}^\beta\right], \quad \abs{\v x - \v x^\beta} > a\\
    &=-\frac{\chi}{8\pi\mu} \sum_{\beta=1,\beta\neq\alpha}^N \left\{\mathcal{F}\left[J_{ij}^o(\v x-\v x^\beta)\right] F_j^\beta - \mathcal{T}\left[J_{ij}^o(\v x-\v x^\beta)\right] T_j^\beta - \mathcal{S}\left[J_{ij}^o(\v x-\v x^\beta)\right] S_{jk}^\beta\right\}
\end{aligned}
    \label{eqn-ambient-flow-diff-odd}
\end{equation}
where the approximation sign is due to the possibly of many-body effects, and that the above results are exact only for an infinitely dilute system. The differential form of linear response functions $\mathcal{F}\left[J^o_{ij}(\v x-\v y)\right]$, $\mathcal{T}\left[J^o_{ij}(\v x-\v y)\right]$ and $\mathcal{S}\left[J^o_{ij}(\v x-\v y)\right]$ can therefore be extracted as: 
\begin{subequations}
\begin{align}
    \mathcal{F}\left[J_{ij}^o(\v x-\v y)\right]&=\left(1+\frac{a^2}{6}\nabla^2\right)J_{ij}^o(\v x - \v y),\quad\abs{\v x - \v y} > 2a\\
    \mathcal{T}\left[J_{ij}^o(\v x-\v y)\right]&=\left(1+\frac{a^2}{10}\nabla^2\right)R_{ij}^o(\v x - \v y),\quad\abs{\v x - \v y} > 2a\\
    \mathcal{S}\left[J_{ij}^o(\v x-\v y)\right]&=\left(1+\frac{a^2}{10}\nabla^2\right)K_{ijk}^o(\v x - \v y),\quad\abs{\v x - \v y} > 2a
\end{align}
\label{eqn-response-diff}
\end{subequations}

\subsubsection{The Fax\'en laws in differential form}
Similarly, with the differential forms of the response functions (Eqn. \ref{eqn-response-diff}) and the generalized Fax\'en laws for odd viscosity (Eqn. \ref{eqn-Faxen-laws-odd}), the differential form of Fax\'en laws with odd viscosity can be written explicitly as
\begin{equation}
\left\{
\begin{aligned}
U_i^\alpha - U_i^\infty(\v x^\alpha) &=- \mathcal{F}\left[u_i^{o\prime}(\v x);\v x^\alpha\right]=-\left(1+\frac{a^2}{6}\nabla^2\right)u_i^{o\prime}(\v x^\alpha)\\
\Omega_i^\alpha - \Omega_i^\infty(\v x^\alpha)&=-\mathcal{T}[u_i^{o\prime}(\v x);\v x^\alpha]=-\left(1+\frac{a^2}{10}\nabla^2\right) \frac{1}{2}\varepsilon_{ijk}\nabla_j u_k^{o\prime}(\v x^\alpha)\\
E_{ij}^\alpha-E_{ij}^\infty(\v x^\alpha)&=-\mathcal{S}[u_i^{o\prime}(\v x);\v x^\alpha]=-\left(1+\frac{a^2}{10}\nabla^2\right)\frac{1}{2}\left[\nabla_ju_i^{o\prime}(\v x^\alpha)+\nabla_iu_j^{o\prime}(\v x^\alpha)\right]
\end{aligned}
\right.
\label{eqn-Faxen-diff-odd}
\end{equation}
where $\mathcal{F}\left[u_i^{o\prime}(\v x);S^\alpha\right]$, $\mathcal{T}\left[u_i^{o\prime}(\v x);S^\alpha\right]$ and $\mathcal{S}\left[u_i^{o\prime}(\v x);S^\alpha\right]$ are the same linear functions defined by Eqn. \ref{eqn-response-diff} but applied to the ambient \emph{odd} flow $\v u^{o\prime}(\v x)$ given by Eqn. \ref{eqn-ambient-flow-diff-odd} and evaluated at the surface of $\alpha$-th particle $S^\alpha$ instead.

\subsubsection{Far-field mobility tensors}
Similarly, with the differential form of Fax\'en laws (Eqn. \ref{eqn-Faxen-diff-odd}) and the expression of ambient flow (Eqn. \ref{eqn-ambient-flow-diff-odd}), the far-field pair mobility tensors $(\alpha\neq\beta)$ can be derived as:
\begin{subequations}
    \begin{align}
    \ten{M}_{UF}^{\alpha\beta}&=\left(M\indices{^o_{UF}^{\alpha\beta}_{ij}}\right)=\frac{\chi}{8\pi\mu}\left(1+\frac{a^2}{6}\nabla^2\right)\left(1+\frac{a^2}{6}\nabla^2\right)J_{ij}^o(\v x^\alpha - \v x^\beta)\\
    \ten{M}_{UT}^{\alpha\beta}&=\left(M\indices{^o_{UT}^{\alpha\beta}_{ij}}\right)=\frac{-\chi}{8\pi\mu}\left(1+\frac{a^2}{6}\nabla^2\right)\left(1+\frac{a^2}{10}\nabla^2\right)R_{ij}^o(\v x^\alpha - \v x^\beta)\\
    \ten{M}_{US}^{\alpha\beta}&=\left(M\indices{^o_{US}^{\alpha\beta}_{ijk}}\right)=\frac{-\chi}{8\pi\mu}\left(1+\frac{a^2}{6}\nabla^2\right)\left(1+\frac{a^2}{10}\nabla^2\right)K_{ijk}^o(\v x^\alpha - \v x^\beta)\\
    \ten{M}_{\Omega T}^{\alpha\beta}&=\left(M\indices{^o_{\Omega T}^{\alpha\beta}_{ij}}\right)=\frac{-\chi}{8\pi\mu}\left(1+\frac{a^2}{10}\nabla^2\right)\left(1+\frac{a^2}{10}\nabla^2\right)\frac{1}{2}\varepsilon_{ikl}\nabla_kR_{lj}^o(\v x^\alpha - \v x^\beta)\\
    \ten{M}_{\Omega S}^{\alpha\beta}&=\left(M\indices{^o_{\Omega S}^{\alpha\beta}_{ijk}}\right)=\frac{-\chi}{8\pi\mu}\left(1+\frac{a^2}{10}\nabla^2\right)\left(1+\frac{a^2}{10}\nabla^2\right)\frac{1}{2}\varepsilon_{ilm}\nabla_lK_{mjk}^o(\v x^\alpha - \v x^\beta)\\
    \ten{M}_{E S}^{\alpha\beta}&=\left(M\indices{^o_{ES}^{\alpha\beta}_{ijkl}}\right)=\frac{-\chi}{8\pi\mu}\left(1+\frac{a^2}{10}\nabla^2\right)\left(1+\frac{a^2}{10}\nabla^2\right) \frac{1}{2}\left[\nabla_j K_{ikl}^o(\v x^\alpha - \v x^\beta)+\nabla_i K_{jkl}^o(\v x^\alpha - \v x^\beta)\right]
    \end{align}
    \label{eqn-mobility-tensors-diff-odd}
\end{subequations}
Note that the odd Oseen tensor is also biharmonic, that is, $\nabla^4 J^o_{ij}(\v r)=0$, which enables further simplification of mobility tensors. The explicit expressions of all far-field odd mobility tensors can be found at Appx. \ref{appx-sd-mobility-far-pair-odd}.

By construction, the near-field mobility tensors (Appx. \ref{appx-sd-mobility-near-pair-odd}) should match with the far-field mobility tensors (Appx. \ref{appx-sd-mobility-far-pair-odd}) continuously at the particle surface. Such a continuity requirement can be checked explicitly by taking limits of both near-field and far-field expressions to the particle surface.

There are also many symmetry relations of mobility tensors of odd viscosity constrained by symmetry properties of systems under consideration. Due to the existence of odd viscosity, some symmetry properties of system are changed comparing even viscosity mobility tensors, which yields different sets of symmetry relations comparing with those of even viscosity. There are symmetry relations of component indices ($i,j,k,l$) (Appx. \ref{appx-sd-symmetry-odd-space}) and particle indices ($\alpha,\beta$) (Appx. \ref{appx-sd-symmetry-odd-parity}), which reflects the space symmetry and the parity symmetry of system, respectively. There are also symmetry relations constrained by the generalized Lorentz reciprocal theorem (Appx. \ref{appx-sd-symmetry-odd-time}), which reflects the time reversal symmetry of system. For a complete list of symmetry relations, please refer to Appx. \ref{appx-sd-symmetry-odd}.

With all these symmetry relations of the mobility tensors, it can be shown that the grand mobility tensor of odd viscosity $\mathcal{M}^o_\infty$ (Eqn. \ref{eqn-mobility-tensors-diff-odd}) is ``symmetric" and only the upper (or lower) half of the matrix needs to be computed as (Appx. \ref{appx-sd-symmetry-odd-def})
\begin{equation}
    M\indices{^o_{\Omega F}^{\alpha\beta}_{ij}}=M\indices{^o_{U T}^{\alpha\beta}_{ij}},\quad
    M\indices{^o_{E F}^{\alpha\beta}_{ijk}}=-M\indices{^o_{U S}^{\alpha\beta}_{kij}},\quad
    M\indices{^o_{E T}^{\alpha\beta}_{ijk}}=-M\indices{^o_{\Omega S}^{\alpha\beta}_{kij}}
\end{equation}

\subsection{Stokesian Dynamics with odd viscosity}\label{sec-mobility-tensors-full}
Finally, with both the near-field (Eqn. \ref{eqn-mobility-tensors-integral-odd}) and far-field (Eqn. \ref{eqn-mobility-tensors-diff-odd}) \emph{odd} mobility tensors constructed, the traditional Stokesian Dynamics formalism can be modified to incorporate the leading order effects due to odd viscosity. Mathematically, the hydrodynamic interactions in Stokesian Dynamics with both even and odd viscosity can be written as
\begin{equation}
    \mathcal{M}=\mathcal{M}^e + \mathcal{M}^o
    \label{eqn-grand-mobility-full-odd}
\end{equation}
where $\mathcal{M}$ is the full grand mobility tensor, which contains hydrodynamic interactions from both even and odd viscosity; $\mathcal{M}^e$ is the grand mobility tensor due to even viscosity, which includes both near-field and far-field even mobility tensors \cite{Brady1988}; $\mathcal{M}^o$ is the grand mobility tensor due to odd viscosity, which includes both near-field (Eqn. \ref{eqn-mobility-tensors-integral-odd} and Appx. \ref{appx-sd-mobility-near-pair-odd}) and far-field (Eqn. \ref{eqn-mobility-tensors-diff-odd} and Appx. \ref{appx-sd-mobility-far-pair-odd}) odd mobility tensors derived in this work.

Due to the lack of lubrication theory with odd viscosity, it is important to be aware that the current near-field mobility tensors are not solved exactly like in the traditional Stokesian Dynamics framework\cite{ONeill1967,Goldman1967,Goldman1967a}. It primarily serves as a convenient regularized near-field mobility tensors for numeric implementations. When the systems of interests are not very dense, the far-field mobility tensors should be able to capture dominant hydrodynamic interactions in the systems. Lubrication effects should be of more importance for dense suspensions, and current near-field mobility tensors should be avoided in such scenarios.

With the explicit expressions of odd mobility tensors, the extra mobility tensors due to odd viscosity can be readily incorporated into any numeric implementation of traditional Stokesian Dynamics, which enables large-scale simulations (but only practical when integrated into a highly-optimized parallel implementation) of collective motions of particles suspended in a fluid medium with both even and odd viscosity.

\section{Discussion and future works}

In this work, we have extended the traditional Stokesian Dynamics simulation framework to further expand its capability to active fluids exhibiting odd viscosity effects. This enables simulations of the collective behaviors of many particles suspended in a fluid medium with both even and odd viscosity, which may help understanding and interpreting recent experimental advances in active systems.

Furthermore, there are still many open questions that need to be addressed in these non-equilibrium active fluid systems. First of all, we only consider a rather simple form of odd viscosity(Eqn. \ref{eqn-viscosity-odd-2d}) under restrictive assumptions. A more general expression of odd viscosity with multiple coefficients do and should exist\cite{Khain2022}. Other coefficients may need to be included depending on the system setup. Besides, for the continual development of the Stokesian Dynamics with odd viscosity, the corresponding wall corrections\cite{Blake1971,Swan2007} and lubrication theory\cite{ONeill1967,Goldman1967,Goldman1967a} are still missing and worth investigating. For describing Brownian motions of particles, it is also very interesting to check whether the fluctuation-dissipation theorem\cite{Chu2019,Burkholder2019,cengio2020fluctuationdissipation,PhysRevLett.107.130601,Junco2018,Sarracino2019,PhysRevX.9.021009} still holds in such active systems.

Experimentally, it is also worth designing active systems\cite{Farhadi2018,Workamp2018} that isolate hydrodynamic interactions from other interactions for a systematic investigation of the effects of odd viscosity. An active control of odd viscosity measurements will enable the exploration and quantification of additional collective behaviors of particles in the presence of odd viscosity.

In terms of numerical technique development, it is helpful to automate the numeric implementation process for the Stokesian Dynamics community. Since Stokesian Dynamics relies on the analytic expressions of mobility tensors, which are ultimately derived from the Green's function of the system, the whole theoretical derivations of Stokesian Dynamics formalism can be automated by a Symbolic Algebra System, such as SymPy. Then, the corresponding analytic expressions of mobility tensors can be generated automatically by simply changing the fundamental Green's function, such as replacing the Oseen tensor with the odd Oseen tensor. Similar to some existing libraries such as lbmpy\cite{Bauer2021a}, those symbolic expressions can be used for the generation of kernels on targeted devices (such as the CUDA code for GPU devices), where automatic code optimization can be applied to have highly-efficient device-specific implementations.

\begin{acknowledgments}
We thank Prof. Paul Chaikin and Dr. Michael Wang for discussing with us their unpublished experiments which motivated this work. We also thank Prof. Vincenzo Vitelli and his group members for carefully reviewing our work and pointing out the limits where our work becomes consistent with his work in reference\cite{Khain2022}. This work was supported as part of the Center for Bio-Inspired Energy Science, an Energy Frontier Research Center funded by the US Department of Energy, Office of Science, Basic Energy Sciences under Award DE-SC0000989.
\end{acknowledgments}

\appendix

\section{Green's function of odd Stokes' equations}\label{appx-Green-func}
\subsection{Green's function in k-space}\label{appx-Green-func-k-space}
In order to obtain the Green's function of odd Stokes' equations (Eqn. \ref{eqn-Stokes}), we need to find its solution in response to a point force (Eqn. \ref{eqn-Stokes-pf}),
\begin{equation*}
    \begin{pmatrix}
    \mu_s&\mu_o&0\\
    -\mu_o&\mu_s&0\\
    0&0&\mu_s
    \end{pmatrix}
    \nabla^2 \v u = \grad p - \v F \delta(\v r)
\end{equation*}
The above equations can be transformed into the k-space by applying the Fourier transformation to both pressure field and flow fields defined as
\begin{equation}
    \v u(\v r) = \mathcal{F}^{-1}\left[\widehat{\v u}(\v k)\right]=\frac{1}{(2\pi)^3}\int_{\mathbb{R}^3} \mathrm{d}\v k \widehat{\v u}(\v k) e^{i\v k \cdot \v r}
    , \quad
    p(\v r) = \mathcal{F}^{-1}\left[\widehat{p}(\v k)\right]=\frac{1}{(2\pi)^3} \int_{\mathbb{R}^3} \mathrm{d}\v k \widehat{p}(\v k) e^{i\v k \cdot \v r}
\end{equation}
where $\mathcal{F}^{-1}$ denotes the inverse Fourier transformation, $\v k$ is the wave vector and $\widehat{p}(\v k)$ and $\widehat{\v u}(\v k)$ are the corresponding pressure and flow fields in the k-space. Plugging above expressions back into the Stokes' equations (Eqn. \ref{eqn-Stokes-pf}) yields 
\begin{equation*}
    -\mu_{ij}k^2 \widehat{u}_j(\v k) = i k_i \widehat{p}(\v k) - F_i
\end{equation*}
where $\mu_{ij}=\mu_s\delta_{ij}+\mu_o \epsilon_{ij}$. Due to the linearity of odd Stokes' equations, both induced pressure and flow fields should be proportional to external forces (Eqn. \ref{eqn-linearity}), whose corresponding k-space expressions are
\begin{equation*}
    \widehat{p}(\v k)=\widehat{P}_j(\v k) F_j, \quad \widehat{\v u}(\v k) = \widehat{G}_{ij}(\v k) F_j
\end{equation*}
where $\widehat{P}_j(\v k)$ and $\widehat{G}_{ij}(\v k)$ denote the Green's functions of pressure field and flow fields in the k-space, respectively. Such linear relations enable the elimination of external forces $F_i$, which simplify Eqn. \ref{eqn-Stokes-pf-kspace} to
\begin{equation}
    -k^2 \mu_{ik} \widehat{G}_{kj}(\v k) = ik_i \widehat{P}_j(\v k) - \delta_{ij}
    \label{eqn-Stokes-k}
\end{equation}
For simplicity, we will restrict our consideration to incompressible fluids. The incompressibility condition ($\div{\v u}=0$) in the k-space reads
\begin{equation}
    \widehat{G}_{ij}(\v k) k_i = 0
\end{equation}
Taking the inner product of Eqn. \ref{eqn-Stokes-k} with $k_i$ yields
\begin{equation*}
    -k^2\mu_{ik} k_i \widehat{G}_{kj}(\v k) = -k^2 \mu_0 \varepsilon_{ik}k_i \widehat{G}_{kj}(\v k) = ik^2 \widehat{P}_j(\v k) - k_j
\end{equation*}
It gives the Green's function of the pressure field in the k-space as
\begin{equation}
    \widehat{P}_j(\v k) = - \frac{ik_j}{k^2} + i \mu_0 \varepsilon_{ik}k_i \widehat{G}_{kj}(\v k)
    \label{eqn-pressure-k}
\end{equation}
where $\mu_{ij}=\mu_s \delta_{ij} + \mu_o \varepsilon_{ij}$ by definition (Eqn. \ref{eqn-viscosity-rank2}) and the second term is the additional pressure contribution due to the odd viscosity. Plugging the above expression back into Eqn. \ref{eqn-Stokes-k} gives that
\begin{equation}
    \begin{pmatrix}
    \mu_s + \mu_o \frac{k_xk_y}{k^2}& \mu_o\left(1-\frac{k_x^2}{k^2}\right)&0\\
    \mu_o(-1+\frac{k_y^2}{k^2})&\mu_s - \mu_o \frac{k_yk_x}{k^2}&0\\
    \mu_o \frac{k_yk_z}{k^2}&-\mu_o\frac{k_xk_z}{k^2}&\mu_s
    \end{pmatrix} \widehat{G}_{kj}(\v k) = - \frac{k_ik_j}{k^4} + \frac{\delta_{ij}}{k^2}
\end{equation}
The right-hand side is exactly the k-space expression of the Oseen tensor\cite{lisicki2013four} and the matrix on the left-hand side can be directly inverted, which gives the k-space expression of the Green's function of the flow field as
\begin{equation}
    \widehat{G}_{ij}(\v k)=\frac{1}{\mu_s}\frac{1}{\Sigma^2(\v k)}
    \begin{pmatrix}
    \frac{k_y^2+k_z^2}{k^4}&-\frac{k_xk_y}{k^4}-\chi\frac{k_z^2}{k^4}&- \frac{k_xk_z}{k^4}+\chi\frac{k_yk_z}{k^4}\\
    -\frac{k_xk_y}{k^4}+\chi\frac{k_z^2}{k^4}&\frac{k_x^2+k_z^2}{k^4}&-\frac{k_yk_z}{k^4}-\chi\frac{k_xk_z}{k^4}\\
    -\frac{k_xk_z}{k^4}-\chi\frac{k_yk_z}{k^4}&-\frac{k_yk_z}{k^4}+\chi\frac{k_xk_z}{k^4}&\frac{k_x^2+k_y^2}{k^4}
    \end{pmatrix}
    \label{eqn-Green-func-k}
\end{equation}
where two auxiliary variables are introduced as
\begin{equation}
    \Sigma^2(\v k) = 1+\chi^2\frac{k_z^2}{k^2}, \quad \chi = \frac{\mu_o}{\mu_s}
\end{equation}
As any second order tensor can be decomposed into symmetric and anti-symmetric parts, the above k-space Green's function can be split as
\begin{subequations}
\begin{equation}
    \widehat{G}_{ij}(\v k) = \widehat{G}_{(ij)}(\v k) + \widehat{G}_{[ij]}(\v k)
\end{equation}
\begin{equation}
    \widehat{G}_{(ij)}(\v k)=\frac{1}{\mu_s}\frac{1}{\Sigma^2(\v k)}\left(-\frac{k_ik_j}{k^4}+\frac{\delta_{ij}}{k^2}\right)
\end{equation}
\begin{equation}
    \widehat{G}_{[ij]}(\v k)=-\frac{1}{\mu_s} \frac{\chi}{\Sigma^2(\v k)}\varepsilon_{ijk}k_k \frac{k_z}{k^4}
\end{equation}
\label{eqn-Green-func-decomposed}
\end{subequations}
where indices in parentheses and brackets indicate symmetrizing and anti-symmetrizing the orginal tensor, respectively.

\subsection{Green's function in real-space}\label{appx-Green-func-real-space}
In principle, the corresponding real-space expressions of Green's functions can be obtained via the inverse Fourier transformation, that is,
\begin{equation}
    G_{ij}(\v r) = \mathcal{F}^{-1}\left[\widehat{G}_{ij}(\v k)\right], \quad P_{i}(\v r) = \mathcal{F}^{-1}\left[\widehat{P}_{i}(\v k)\right]
\end{equation}
However, analytical calculation of above inverse Fourier transformations is not straightforward due to the additional k-dependence introduced by $\Sigma^2(\v k)$ for non-vanishing odd viscosity. For simplicity, we focus on the fluid with a small odd viscosity\cite{Soni2019}, i.e. $\mu_o \ll \mu_s$. Then, the dimensionless viscosity ratio $\chi \ll 1$ serves as a natural expansion parameter, which physically characterizes the activity level of the system. For the weak activity limit, that is, $\chi\to 0$, the inverse of $\Sigma^2(\v k)$ can be expanded as
\begin{equation}
    \frac{1}{\Sigma^2(\v k)} = \frac{1}{1+\chi^2\frac{k_z^2}{k^2}} = \sum_{n=0}^\infty (-1)^n \left(\frac{k_z^2}{k^2}\chi^2\right)^n = 1 - \frac{k_z^2}{k^2}\chi^2 + \mathcal{O}(\chi^4)
\end{equation}
Then, the expansions of the symmetric and anti-symmetric parts of the k-space Green's function (Eqn. \ref{eqn-Green-func-decomposed}) can be written as
\begin{subequations}
\begin{equation}
    \widehat{G}_{(ij)}(\v k)=\frac{1}{\mu_s}\sum_{n=0}^\infty (-1)^n \left(\frac{k_z^2}{k^2}\chi^2\right)^n\left(-\frac{k_ik_j}{k^4}+\frac{\delta_{ij}}{k^2}\right) = \left[\widehat{G}_{(ij)}^{(0)}+\widehat{G}_{(ij)}^{(2)}\right] + \mathcal{O}(\chi^4)
\end{equation}
\begin{equation}
    \widehat{G}_{[ij]}(\v k)=-\frac{\chi}{\mu_s} \sum_{n=0}^\infty (-1)^n \left(\frac{k_z^2}{k^2}\chi^2\right)^n\left(\varepsilon_{ijk}k_k \frac{k_z}{k^4}\right)= \left[\widehat{G}_{[ij]}^{(1)}+\widehat{G}_{[ij]}^{(3)}\right] + \mathcal{O}(\chi^5)
\end{equation}
\end{subequations}
where $n$-th order expansion of the symmetric and anti-symmetric parts are defined as
\begin{subequations}
\begin{equation}
    \widehat{G}_{(ij)}^{(2n)}(\v k) = \frac{(-1)^n}{\mu_s} \left(\frac{k_z^2}{k^2}\chi^2\right)^n \left(-\frac{k_ik_j}{k^4}+\frac{\delta_{ij}}{k^2}\right), \quad n \in \mathbb{N}
\end{equation}
\begin{equation}
    \widehat{G}_{[ij]}^{(2n+1)}(\v k) =-\chi \frac{(-1)^n}{\mu_s} \left(\frac{k_z^2}{k^2}\chi^2\right)^n \left(\varepsilon_{ijk} \frac{k_kk_z}{k^4}\right), \quad n \in \mathbb{N}
\end{equation}
\end{subequations}
where $\mathbb{N}$ is the set of natural numbers. The inverse Fourier transformation of the lowest order expressions is easy to compute
\begin{subequations}
\begin{equation}
    G_{(ij)}^{(0)}(\v r) = \mathcal{F}^{-1}\left[\widehat{G}_{(ij)}^{(0)}(\v k)\right]=\frac{1}{\mu_s}\mathcal{F}^{-1}\left[-\frac{k_ik_j}{k^4}+\frac{\delta_{ij}}{k^2}\right]=\frac{1}{8\pi\mu_s} \left(\frac{\delta_{ij}}{r}+\frac{r_ir_j}{r^3}\right)
\end{equation}
\begin{equation}
    G_{[ij]}^{(1)}(\v r) = \mathcal{F}^{-1}\left[\widehat{G}_{[ij]}^{(1)}(\v k)\right]=-\frac{1}{\mu_s}\mathcal{F}^{-1}\left[\varepsilon_{ijk} \frac{k_kk_z}{k^4}\right]=-\frac{\chi}{8\pi\mu_s}\varepsilon_{ijk}\left(\frac{\delta_{k3}}{r}-\frac{r_kr_3}{r^3}\right)
\end{equation}
\end{subequations}
Note the expression of the conventional Oseen tensor is
\begin{equation}
    J_{ij}(\v r)=\frac{\delta_{ij}}{r}+\frac{r_ir_j}{r^3}
    \label{eqn-Oseen}
\end{equation}
which enables rewriting the above equations as
\begin{subequations}
\begin{equation}
    G_{(ij)}^{(0)}(\v r) = \frac{1}{8\pi\mu_s} J_{ij}(\v r)
    \label{eqn-Green-func-sym-0}
\end{equation}
\begin{equation}
    G_{[ij]}^{(1)}(\v r) = -\frac{\chi}{8\pi\mu_s}\varepsilon_{ijk}\delta_{l3}\left(\frac{\delta_{kl}}{r}-\frac{r_kr_l}{r^3}\right)=-\frac{\chi}{8\pi\mu_s}\varepsilon_{ijk}\delta_{l3} J_{kl}^s(\v r)=-\frac{\chi}{8\pi\mu_s}J_{ij}^o(\v r)
    \label{eqn-Green-func-antisym-1}
\end{equation}
\end{subequations}
where an \emph{odd} Oseen tensor is analogously defined as
\begin{equation}
    J_{ij}^o(\v r) = \varepsilon_{ijk}\delta_{l3} J_{kl}^s(\v r)=\varepsilon_{ijk}\delta_{l3}\left(\frac{\delta_{kl}}{r}-\frac{r_kr_l}{r^3}\right)
    \label{eqn-Oseen-odd}
\end{equation}
For convenience, an auxiliary symmetric tensor $J_{ij}^s(\v r)$ is also introduced, which only differs with the symmetric Oseen tensor $J_{ij}(\v r)$ by a minus sign of the anisotropic term
\begin{equation}
    J_{ij}^s(\v r) =\frac{\delta_{ij}}{r}-\frac{r_ir_j}{r^3}
    \label{eqn-Oseen-sym}
\end{equation}

As for the higher order terms, it can be constructed from the lower order solutions by noting the following properties of the Fourier transformation
\begin{equation}
    \pdd{}{z} f(\v r) = \frac{1}{(2\pi)^3}\int \mathrm{d}\v k (-k_z^2) \widehat{f}(\v k) e^{i \v k \cdot \v r}
    , \quad
    \nabla^2 f(\v r) = \frac{1}{(2\pi)^3}\int \mathrm{d}\v k (-k^2) \widehat{f}(\v k) e^{i \v k \cdot \v r}
\end{equation}
Therefore, the next order solutions can be constructed as
\begin{subequations}
\begin{equation}
    G_{(ij)}^{(2)}(\v r) =-\chi^2 \mathcal{F}^{-1}\left[\frac{k_z^2}{k^2}\widehat{G}_{(ij)}^{(0)}(\v k)\right]=-\chi^2 \pdd{}{z}\nabla^{-2}\left[ \mathcal{F}^{-1}\left[\widehat{G}_{(ij)}^{(0)}(\v k)\right]\right]
\end{equation}
\begin{equation}
    G_{[ij]}^{(3)}(\v r) =-\chi^2 \mathcal{F}^{-1}\left[\frac{k_z^2}{k^2}\widehat{G}_{[ij]}^{(1)}(\v k)\right]=-\chi^2 \pdd{}{z}\nabla^{-2}\left[ \mathcal{F}^{-1}\left[\widehat{G}_{[ij])}^{(1)}(\v k)\right]\right]
\end{equation}
\end{subequations}
where $\nabla^{-2}$ is the inverse Laplace operator defined as
\begin{equation}
    \nabla^{-2} = \int_{\mathbb{R}^3} \mathrm{d} \v r^\prime G(\v r - \v r^\prime),\quad G(\v r) = -\frac{1}{4\pi r}
\end{equation}
Here, $G(\v r)$ is the Green's function of the 3D Laplace's equation. Also please be aware that the inverse Laplace operator $\nabla^{-2}$ is spherically symmetric but the double derivative in z-direction $\pdd{}{z}$ is not. Thus, it's important to choose a proper order between two operators to maintain the desired symmetry for intermediate steps.

In the next, both symmetric parts $\left( G_{(ij)}^{(2)}(\v r)\right)$ and anti-symmetric parts $\left( G_{[ij]}^{(3)}(\v r)\right)$ will be calculated as an illustration and higher order terms can be computed in a similar manner.
\begin{itemize}
    \item the symmetric part $G_{(ij)}^{(2)}(\v r)$\\
    For the symmetric part, the inverse Laplace operator can be computed as:
    \begin{equation}
        \nabla^{-2}\left[\mathcal{F}^{-1}\left[G_{(ij)}^{(0)}\right]\right]=\frac{1}{8\pi\mu_s}\nabla^{-2} \left[\frac{\delta_{ij}}{r}+\frac{r_ir_j}{r^3}\right]
        =\frac{1}{8\pi\mu_s} \int_{\mathbb{R}^3} \mathrm{d} \v r' G(\v r - \v r') J_{ij}(\v r')
    \end{equation}
    The inverse Laplace operator of the Oseen tensor can be evaluated in spherical coordinates, which gives
    \begin{equation}
        \nabla^{-2}\left[J_{ij}(\v r)\right]=\frac{3}{4}r\delta_{ij} - \frac{1}{4}\frac{r_ir_j}{r}
    \end{equation}
    Then, the double derivative in z-direction is straightforward to calculate, which gives the real-space expression of the symmetric part as
    \begin{equation}
    \begin{aligned}
        G_{(ij)}^{(2)}(\v r) &= - \frac{\chi^2}{8\pi\mu_s} \pdd{}{z} \left(\frac{3}{4}r\delta_{ij} - \frac{1}{4}\frac{r_ir_j}{r}\right)\\
        &=-\frac{\chi^2}{8\pi\mu_s}\left[\frac{3}{4}\left(1-\frac{r_3^2}{r^2}\right)\frac{1}{r}\delta_{ij}+\frac{1}{4}\left(1-3\frac{z^2}{r^2}\right)\frac{r_ir_j}{r^3}\right.\\
        &\left.+\frac{1}{2}\frac{r_3}{r^3}\left(\delta_{i3}r_j+r_i\delta_{j3}\right)-\frac{1}{2}\frac{1}{r}\delta_{i3}\delta_{j3}\right]
        \label{eqn-Green-func-sym-2}
    \end{aligned}
    \end{equation}
    \item the anti-symmetric part $G_{[ij]}^{(3)}(\v r)$\\
    For the anti-symmetric part, the inverse Laplace operator can be calculated as
    \begin{equation}
    \begin{aligned}
        \nabla^{-2}\left[\mathcal{F}^{-1}\left[G_{[ij]}^{(1)}\right]\right]&=-\frac{\chi}{8\pi\mu_s}\varepsilon_{ijk}\delta_{l3}\nabla^{-2} \left[\frac{\delta_{kl}}{r}-\frac{r_kr_l}{r^3}\right]\\
        &=-\frac{\chi}{8\pi\mu_s}\varepsilon_{ijk}\delta_{l3}\nabla^{-2} \int_{\mathbb{R}^3} \mathrm{d} \v r' G(\v r - \v r') J^s_{ij}(\v r')
    \end{aligned}
    \end{equation}
    The inverse Laplace operator of $J^s_{ij}(\v r)$ can be similarly evaluated in the spherical coordinates, which gives
    \begin{equation}
        \nabla^{-2}\left[J^s_{ij}(\v r)\right]=\frac{1}{4}r\delta_{ij}+\frac{1}{4}\frac{r_ir_j}{r}
    \end{equation}
    Then, plugging it back and evaluate the double derivative in z-direction yields the real-space expression of the anti-symmetric part as
    \begin{equation}
    \begin{aligned}
        G_{[ij]}^{(3)}(\v r)&=\frac{\chi^3}{8\pi\mu_s}\varepsilon_{ijk}\delta_{l3} \pdd{}{z}\left(\frac{1}{4}r\delta_{kl}+\frac{1}{4}\frac{r_kr_l}{r}\right)\\
        &=\frac{\chi^3}{8\pi\mu_s}\varepsilon_{ijk}\delta_{l3}\left[\frac{1}{4}\left(1-\frac{r_3^2}{r^2}\right)\frac{1}{r}\delta_{kl}-\frac{1}{4}\left(1-3\frac{z^2}{r^2}\right)\frac{r_kr_l}{r^3}\right.\\
        &\left.-\frac{1}{2}\frac{r_3}{r^3}\left(\delta_{k3}r_l+r_k\delta_{l3}\right)+\frac{1}{2}\frac{1}{r}\delta_{k3}\delta_{l3}\right]\\
        &=\frac{\chi^3}{8\pi\mu_s}\left[\frac{3}{4}\left(1-\frac{r_3^2}{r^2}\right)\frac{1}{r}\varepsilon_{ij3} - \frac{3}{4}\left(1-\frac{r_3^2}{r^2}\right)\frac{r_3}{r^3}\varepsilon_{ijk}r_k\right]
    \end{aligned}
    \label{eqn-Green-func-antisym-3}
    \end{equation}
\end{itemize}
Finally, the Green's function in the real-space can be expressed as
\begin{subequations}
\begin{equation}
    G_{ij}(\v r) = G_{(ij)}(\v r) + G_{[ij]}(\v r)
\end{equation}
\begin{equation}
    G_{(ij)}(\v r) = G_{(ij)}^{(0)}(\v r) + G_{(ij)}^{(2)}(\v r) + \mathcal{O}(\chi^4)
\end{equation}
\begin{equation}
    G_{[ij]}(\v r) = G_{[ij]}^{(1)}(\v r) + G_{[ij]}^{(3)}(\v r) + \mathcal{O}(\chi^5)
\end{equation}
\label{eqn-Green-func-real}
\end{subequations}
where the explicit expressions of the symmetric Green's functions are given by Eqn. \ref{eqn-Green-func-sym-0} and \ref{eqn-Green-func-sym-2}; the explicit expressions of the anti-symmetric Green's functions are given by Eqn. \ref{eqn-Green-func-antisym-1} and \ref{eqn-Green-func-antisym-3}.

\section{Generalized Lorentz reciprocal theorem}\label{appx-lrt}
Suppose that $(\v u^{(1)},\gv \sigma^{(1)})$ and $(\v u^{(2)}, \gv \sigma^{(2)})$ are two different solutions of the \emph{odd} Stokes' equations with the same fluid domain $V$ but subjected to different boundary conditions at its bounding surface $S$. Due to the linearity of the viscosity tensor, the overall stress tensor can be separated into stresses contributed by even and odd viscosity, that is, $\ten{\sigma}=\ten{\sigma}_e + \ten{\sigma}_o$. Then, the generalized Lorentz reciprocal theorem\index{Lorentz reciprocal theorem} states that
\begin{equation}
\begin{aligned}
    &\oint_{S} \v u^{(1)} \cdot \left[\left(\gv \sigma^{(2)}_e + \gv \sigma^{(2)}_o \right)\cdot \uv n\right] \mathrm{d}S - \int_{V} \v u^{(1)} \cdot \left[\div{\left(\gv \sigma^{(2)}_e + \gv \sigma^{(2)}_o \right)}\right]\mathrm{d}V \\
    &= \oint_{S} \v u^{(2)}\cdot \left[\left(\gv \sigma^{(1)}_e - \gv \sigma^{(1)}_o \right) \cdot \uv n\right] \mathrm{d}S - \int_{V} \v u^{(2)} \cdot \left[\div{\left(\gv \sigma^{(1)}_e - \gv \sigma^{(1)}_o \right)}\right]\mathrm{d}V
\end{aligned}
\label{eqn-LRT}
\end{equation}
The proof of the above generalized Lorentz reciprocal theorem is analogous to the proof of the conventional Lorentz reciprocal theorem with only even viscosity\cite{Graham2018,Guazzelli2011,SangtaeKim2005}. By applying the divergence theorem, it is equivalent to prove that
\begin{equation}
\begin{aligned}
    &\nabla_i \left[u^{(1)}_j \sigma^{(2)}_{e\ ij} + u^{(1)}_j \sigma^{(2)}_{o\ ij}\right] - u^{(1)}_i \nabla_j \left[\sigma_{e\ ij}^{(2)} + \sigma_{o\ ij}^{(2)}\right] \\
    &= \nabla_i \left[u^{(2)}_j \sigma^{(1)}_{e\ ij} - u^{(2)}_j \sigma^{(1)}_{o\ ij}\right] - u^{(2)}_i \nabla_j \left[ \sigma_{e\ ij}^{(1)} - \sigma_{o\ ij}^{(1)}\right]
\end{aligned}
\end{equation}
Note the following useful identities
\begin{subequations}
\begin{equation}
    \sigma^{(2)}_{ij} \nabla_j u_i^{(1)}=\nabla_j\left(\sigma_{ij}^{(2)}u_i^{(1)}\right) - u_i^{(1)} \nabla_j \sigma_{ij}^{(2)}=\nabla_i\left(u_j^{(1)}\sigma_{ij}^{(2)}\right) - u_i^{(1)} \nabla_j \sigma_{ij}^{(2)}
\end{equation}
\begin{equation}
    \sigma^{(1)}_{ij} \nabla_j u_i^{(2)}=\nabla_j\left(\sigma_{ij}^{(1)}u_i^{(2)}\right) - u_i^{(2)} \nabla_j \sigma_{ij}^{(1)}=\nabla_i\left(u_j^{(2)}\sigma_{ij}^{(1)}\right) - u_i^{(2)} \nabla_j \sigma_{ij}^{(1)}
\end{equation}
\end{subequations}
where stress tensors are assumed to be symmetric, i.e. $\sigma_{ij}^{(1)}=\sigma_{ji}^{(1)}$ and $\sigma_{ij}^{(2)}=\sigma_{ji}^{(2)}$. This holds for both even and odd viscous stress tensors as long as there is no external angular momentum in the system. Therefore, it further reduces to prove that
\begin{equation}
    \sigma_{e\ ij}^{(2)} \nabla_j u_i^{(1)} + \sigma_{o\ ij}^{(2)} \nabla_j u_i^{(1)} = \sigma_{e\ ij}^{(1)} \nabla_j u_i^{(2)} - \sigma_{o\ ij}^{(1)} \nabla_j u_i^{(2)}
\end{equation}
For a general linear constitutive relation of Newtonian liquids, the relation between the stress tensor and the velocity gradient tensor can be written as
\begin{equation}
    \sigma_{ij}^{(1)} = \sigma_{e\ ij}^{(1)}+\sigma_{o\ ij}^{(1)}=\left(\mu^e_{ijkl}+\mu^o_{ijkl}\right) \nabla_l u_k^{(1)},\quad \sigma_{ij}^{(2)} = \sigma_{e\ ij}^{(2)} + \sigma_{o\ ij}^{(2)} = \left(\mu^e_{ijkl}+\mu^o_{ijkl}\right) \nabla_l u_k^{(2)}
\end{equation}
Then, it's equivalent to prove that
\begin{equation}
    \mu_{ijkl}^e \nabla_l u_k^{(2)} \nabla_j u_i^{(1)} + \mu_{ijkl}^o \nabla_l u_k^{(2)} \nabla_j u_i^{(1)} =
    \mu_{ijkl}^e \nabla_l u_k^{(1)} \nabla_j u_i^{(2)} - \mu_{ijkl}^o \nabla_l u_k^{(1)} \nabla_j u_i^{(2)}
\end{equation}
For the above equation to hold everywhere in the fluid domain, it requires that
\begin{equation}
    \mu^e_{ijkl} = \mu^e_{klij}, \quad \mu^o_{ijkl} =-\mu^o_{klij}
\end{equation}
These symmetry relations correspond to the time reversal symmetry of the system and are fulfilled by even and odd viscosity tensors respectively (Eqn. \ref{eqn-viscosity-even} and \ref{eqn-viscosity-odd}), which completes the proof of the generalized Lorentz reciprocal theorem.

\section{Generalized Fax\'en laws}\label{appx-faxen-laws}
The Fax\'en laws connect the kinetic motions $(\v U^\alpha, \v \Omega^\alpha, \ten{E}^\alpha)$ with corresponding mechanical quantities $(\v F^\beta, \v T^\beta, \ten{S}^\beta)$ of each particle. However, the validity of the Fax\'en laws with even viscosity relies on the Lorentz reciprocal theorem and symmetry properties of the Oseen tensor\cite{SangtaeKim2005}. With the generalized Lorentz reciprocal theorem derived in Appx. \ref{appx-lrt}, the conventional Fax\'en laws will also be generalized here to include additional effects due to odd viscosity.

Following the proof of conventional Fax\'en laws described in \cite{SangtaeKim2005}, let us consider a particle that translates in the fluid with velocity $\v U_1$, which has a corresponding flow field $\v u_1$, and another particle that is stationary but that there is a point-force $\v F$ applied at location $\v y$ outside the particle. Then, applying the generalized Lorentz theorem gives 
\begin{equation}
    \v U_1 \cdot \v F_2 - \v u_1(\v y) \cdot \v F = 0
\end{equation}
where $\v F_2$ is the force acting on the stationary particle due to the corresponding flow field $\v u_2(\v x)$ induced by the point force $\v F$. Because of the linearity of the \emph{odd} Stokes' equations (Eqn. \ref{eqn-Stokes}), the ambient flow for the translating particle can be expressed as
\begin{equation}
    \v u_1(\v x) = \left\{ \mathcal{F}_e\left[\frac{J_{ij}(\v x - \gv \xi) }{8\pi\mu}\right] + \mathcal{F}_o\left[\frac{- \chi J^o_{ij}(\v x - \gv \xi)}{8\pi\mu} \right]\right\}\cdot \v U_1
\end{equation}
where $\mathcal{F}_e$ and $\mathcal{F}_o$ represent a linear response function due to even and odd viscosity, respectively; $\mathcal{F}_e$ and $\mathcal{F}_o$ may or may not have the same functional form. Here, $\gv \xi$ denotes the region where the source force densities are distributed. Plugging it back enables the elimination of $\v U_1$, which yields
\begin{equation}
    F^2_i = \left\{ \mathcal{F}_e\left[\frac{J_{ji}(\v x - \gv \xi) }{8\pi\mu}\right] + \mathcal{F}_o\left[\frac{- \chi J^o_{ji}(\v x - \gv \xi)}{8\pi\mu} \right]\right\}F_j
\end{equation}
Note the symmetry properties of both even and odd Oseen tensors
\begin{subequations}
\begin{equation}
    J_{ji}(\v x - \gv \xi) = J_{ij}(\v x - \gv \xi), \quad J^o_{ji}(\v x - \gv \xi) = -J^o_{ij}(\v x - \gv \xi)
\end{equation}
\begin{equation}
    J_{ij}(\v x - \gv \xi) = J_{ij}(\gv \xi - \v x), \quad J^o_{ij}(\v x - \gv \xi) = J^o_{ij}(\gv \xi - \v x)
\end{equation}
\end{subequations}
Applying those symmetry properties gives
\begin{equation}
    F^2_i =\mathcal{F}_e\left[\frac{J_{ij}(\gv \xi - \v y) }{8\pi\mu}F_j\right]+\mathcal{F}_o\left[\chi\frac{J^o_{ij}(\gv \xi - \v y) }{8\pi\mu}F_j\right]
\end{equation}
On the other hand, the flow fields contributed by even and odd viscosity induced by a point force $\v F$ are:
\begin{equation}
    \v u^e(\gv \xi) = \frac{J_{ij}(\gv \xi - \v y)}{8\pi\mu}F_j
    , \quad
    \v u^o(\gv \xi) =-\chi \frac{J^o_{ij}(\gv \xi - \v y)}{8\pi\mu}F_j
\end{equation}
where $\v u^e(\gv \xi)$ and $\v u^o(\gv \xi)$ denote the flow contributed by the even and odd viscosity, respectively. Thus, the terms inside the brackets are identified as just the induced ambient flow, which gives
\begin{equation}
    \v F_2 =\mathcal{F}_e\left[\v u^e(\gv \xi)\right]-\mathcal{F}_o\left[\v u^o(\gv \xi)\right]
    \label{eqn-faxen-force}
\end{equation}
It shows that there is just one additional minus sign for the flow contributed the odd viscosity in the generalized Fax\'en laws. Following the above construction, one can prove similar relations for torques and stresslets, that is,
\begin{equation}
    \v T = \mathcal{T}_e\left[\v u^e(\gv \xi)\right]-\mathcal{T}_o\left[\v u^o(\gv \xi)\right]
    \label{eqn-faxen-torque}
\end{equation}
\begin{equation}
    \ten S = \mathcal{S}_e\left[\v u^e(\gv \xi)\right]-\mathcal{S}_o\left[\v u^o(\gv \xi)\right]
    \label{eqn-faxen-stresslet}
\end{equation}

In this work, we will only consider additional new contributions due to the odd viscosity in the Fax\'en laws. The Fax\'en laws for even viscosity can be easily found in most micro-hydrodynamics books\cite{SangtaeKim2005,Guazzelli2011,Graham2018}. For a collection of particles, the linear superposition enables writing the ambient \emph{odd} flow field $\v u^{o\prime}$ induced by all other particles excluding the $\alpha$-th particle itself as
\begin{equation}
\begin{aligned}
    u_i^{o\prime} (\v x)
    &\approx-\frac{\chi}{8\pi\mu} \sum_{\beta=1,\beta\neq\alpha}^N \left\{\mathcal{F}\left[J^o_{ij}(\v x-\v x^\beta)\right] F_j^\beta + \mathcal{T}\left[J^o_{ij}(\v x-\v x^\beta)\right] T_j^\beta + \mathcal{S}\left[J^o_{ij}(\v x-\v x^\beta)\right] S_{jk}^\beta\right\}
\end{aligned}
\end{equation}
where $\mathcal{F}\left[J^o_{ij}(\v x-\v y)\right]$, $\mathcal{T}\left[J^o_{ij}(\v x-\v y)\right]$ and $\mathcal{S}\left[J^o_{ij}(\v x-\v y)\right]$ denote linear functions which describe the odd viscosity related responses to external forces, torques, stresslets respectively. Based on the results derived in Eqn. \ref{eqn-faxen-force}, \ref{eqn-faxen-torque} and \ref{eqn-faxen-stresslet}, 
the kinetic motions of $\alpha$-th particle due to the ambient \emph{odd} flow $\v u^{o\prime}$ induced by other particles ($\beta\neq\alpha$) can be therefore written as
\begin{equation}
\left\{
\begin{aligned}
U_i^\alpha - U_i^\infty(\v x^\alpha) &=- \mathcal{F}\left[u_i^{o\prime}(\v x);\v x^\alpha\right]\\
\Omega_i^\alpha - \Omega_i^\infty(\v x^\alpha)&=-\mathcal{T}[u_i^{o\prime}(\v x);\v x^\alpha]\\
E_{ij}^\alpha-E_{ij}^\infty(\v x^\alpha)&=-\mathcal{S}[u_i^{o\prime}(\v x);\v x^\alpha]
\end{aligned}
\right.
\label{eqn-Faxen-laws-odd}
\end{equation}
The above relations are the Fax\'en laws for the odd viscosity, which differ with the Fax\'en laws for the even viscosity by a minus sign.

\section{Explicit expressions of the odd mobility tensors}
In this section, the explicit expressions of odd mobility tensors are documented for reference. All $N$ particles are assumed to be equal size of radius $a$, i.e. $a^\alpha=a^\beta=a$. All distance quantities are normalized by the particle radius, such as $r=\abs{\v x^\alpha - \v x^\beta}/a$. All mobility expressions are normalized by $8\pi\mu a^n$, where the exponent $n$ is chosen to make the corresponding mobility tensor dimensionless, i.e. $\hat{M}\indices{^o_{XY}^{\alpha\beta}_{ij}} = 8\pi\mu a^n M\indices{^o_{XY}^{\alpha\beta}_{ij}}$. The hat symbol $\ \hat{\ }\ $ indicates the quantity is dimensionless and the additional superscript $^o$ means it's an odd viscosity related quantity; The subscript $XY$ means the mobility tensor describes the coupling between kinetic motions $X$ ($X=U$, $\Omega$ or $E$) and force moments $Y$ ($Y=F$, $T$ or $S$); The superscript $\alpha\beta$ specifies the mobility tensor describes the interaction between $\alpha$-th and $\beta$-th particles ($\alpha,\beta=1\cdots N$); The subscript $ijkl$ indicates the Cartesian components of the mobility tensor ($i,j,k,l=1,2,3$). $\delta_{ij}$ is the Kronecker delta and $\varepsilon_{ijk}$ is the right-handed Levi-Civita symbol. Only six out of nine mobility tensors are given and the rest can be obtained by the symmetry relations of mobility tensors. Only odd mobility tensors in the first order of the dimensionless viscosity ratio $\chi$ (odd viscosity $\mu_o$ over even viscosity $\mu_s$) are derived. If necessary, higher order corrections can be computed in a similar manner.
\subsection{Self-mobility tensors}\label{appx-sd-mobility-self-odd}
The dimensionless expressions of mobility tensors for a single isolated particle ($\alpha=\beta$) are shown below
\begin{subequations}
\begin{align}
    \hat{M}\indices{^o_{UF}^{\alpha\alpha}_{ij}}&=\frac{2\chi}{3}\varepsilon_{ij3}\\
    \hat{M}\indices{^o_{UT}^{\alpha\alpha}_{ij}}&=0\\
    \hat{M}\indices{^o_{US}^{\alpha\alpha}_{ijk}}&=0\\
    \hat{M}\indices{^o_{\Omega T}^{\alpha\alpha}_{ij}}&=\frac{\chi}{2}\varepsilon_{ij3}\\
    \hat{M}\indices{^o_{\Omega S}^{\alpha\alpha}_{ijk}}&=\frac{3\chi}{5}\delta_{3ijk}\\
    \hat{M}\indices{^o_{ES}^{\alpha\alpha}_{ijkl}}&=\frac{\chi}{10}\left[\left(\varepsilon_{ik3}\delta_{lj}+\varepsilon_{il3}\delta_{kj}\right) + \left(\varepsilon_{jk3}\delta_{li}+\varepsilon_{jl3}\delta_{ki}\right)\right]
\end{align}
\end{subequations}

\subsection{Pair-mobility tensors ($r > 2a$)}\label{appx-sd-mobility-far-pair-odd}
The definitions of far-field mobility tensors between two different non-overlapping particles ($\alpha\neq\beta$ and $r>2a$) are given at Eqn. \ref{eqn-mobility-tensors-diff-odd} and its corresponding dimensionless explicit expressions are shown below\\
\begin{subequations}
\begin{align}
    \hat{M}\indices{^o_{UF}^{\alpha\beta}_{ij}}&=\frac{\chi}{r}\left[\left(1-\frac{2}{3r^2}\right)\varepsilon_{ij3}-\left(1-\frac{2}{r^2}\right)\frac{\varepsilon_{ijk}r_k r_3}{r^2}\right]\\
    \hat{M}\indices{^o_{UT}^{\alpha\beta}_{ij}}&=-\frac{\chi}{2r^2}\left[\left(1+\frac{8}{5}\frac{1}{r^2}\right)\delta_{ij}\frac{r_3}{r}-\left(1-\frac{8}{5}\frac{1}{r^2}\right)\left(\delta_{i3}\frac{r_j}{r}+\delta_{j3}\frac{r_i}{r}\right)\right]\\
    &\notag-\frac{3\chi}{2r^2}\left(1-\frac{8}{3}\frac{1}{r^2}\right)\frac{r_ir_jr_3}{r^3}\\
    \hat{M}\indices{^o_{US}^{\alpha\beta}_{ijk}}&=\frac{\chi}{2r^2}\left(1-\frac{8}{5}\frac{1}{r^2}\right)\left[\left(\varepsilon_{ij3}\frac{r_k}{r}+\varepsilon_{ik3}\frac{r_j}{r}\right)+\left(\varepsilon_{ijm}\delta_{k3}+\varepsilon_{ikm}\delta_{j3}\right)\frac{r_m}{r}\right]\\
    &\notag-\frac{3\chi}{2r^2}\left(1-\frac{8}{3}\frac{1}{r^2}\right)\left(\varepsilon_{ijm}\frac{r_k}{r}+\varepsilon_{ikm}\frac{r_j}{r}\right)\frac{r_mr_3}{r^2}\\
    \hat{M}\indices{^o_{\Omega T}^{\alpha\beta}_{ij}}&=\frac{\chi}{2r^3}\varepsilon_{ij3} - \frac{3\chi}{2r^3}\varepsilon_{ijl}\frac{r_lr_3}{r^2}\\
    \hat{M}\indices{^o_{\Omega S}^{\alpha\beta}_{ijk}}&=-\frac{\chi}{2r^3}\left(1-\frac{6}{5}\frac{1}{r^2}\right)\delta_{i3}\delta_{jk}+\frac{3\chi}{5r^5}\left(\delta_{ij}\delta_{k3}+\delta_{ik}\delta_{j3}\right)\\
    &\notag-\frac{\chi}{2r^3}\left(15-\frac{42}{r^2}\right)\frac{r_ir_jr_kr_3}{r^4}-\frac{3\chi}{r^5}\left(\delta_{ik}\frac{r_j}{r}+\delta_{ij}\frac{r_k}{r}\right)\frac{r_3}{r}\\
    &\notag+\frac{3\chi}{2r^3}\left(1-\frac{2}{r^2}\right)\left(\delta_{i3}\frac{r_jr_k}{r^2}+\delta_{j3}\frac{r_ir_k}{r^2}+\delta_{k3}\frac{r_ir_j}{r^2}+\delta_{jk}\frac{r_ir_3}{r^2}\right)\\
    \hat{M}\indices{^o_{E S}^{\alpha\beta}_{ijkl}}
    &=\frac{\chi}{4r^3}\left(1-\frac{6}{5}\frac{1}{r^2}\right)\left[\left(\varepsilon_{ik3}\delta_{lj}+\varepsilon_{il3}\delta_{kj}\right) + \left(\varepsilon_{jk3}\delta_{li}+\varepsilon_{jl3}\delta_{ki}\right)\right]\\
    &\notag\frac{15\chi}{4r^3}\left(1-\frac{14}{5r^2}\right)\left[\left(\varepsilon_{ikn}\frac{r_l}{r} + \varepsilon_{iln}\frac{r_k}{r}\right)\frac{r_j}{r}+\left(\varepsilon_{jkn}\frac{r_l}{r} + \varepsilon_{jln}\frac{r_k}{r}\right)\frac{r_i}{r}\right]\frac{r_nr_3}{r^2}\\
    &\notag-\frac{3\chi}{4r^3}\left(1-\frac{2}{r^2}\right)\left[\left(\varepsilon_{ik3}\frac{r_l}{r} + \varepsilon_{il3}\frac{r_k}{r}\right)\frac{r_j}{r} + \left(\varepsilon_{jk3}\frac{r_l}{r} + \varepsilon_{jl3}\frac{r_k}{r}\right)\frac{r_i}{r} \right.\\
    &\notag+\varepsilon_{ikn}\frac{r_n}{r}\left(\delta_{l3}\frac{r_j}{r} + \delta_{lj}\frac{r_3}{r} + \delta_{j3}\frac{r_l}{r}\right)+\varepsilon_{iln}\frac{r_n}{r}\left(\delta_{k3}\frac{r_j}{r} + \delta_{kj}\frac{r_3}{r} + \delta_{j3}\frac{r_k}{r}\right)\\
    &\notag+\left.\varepsilon_{jkn}\frac{r_n}{r}\left(\delta_{l3}\frac{r_i}{r} + \delta_{li}\frac{r_3}{r} + \delta_{i3}\frac{r_l}{r}\right)+\varepsilon_{jln}\frac{r_n}{r}\left(\delta_{k3}\frac{r_i}{r} + \delta_{ki}\frac{r_3}{r} + \delta_{i3}\frac{r_k}{r}\right)\right]
\end{align}
\end{subequations}

\subsection{Pair-mobility tensors ($r \le 2a$)}\label{appx-sd-mobility-near-pair-odd}
The definitions of near-field mobility tensors between two different overlapping particles ($\alpha\neq\beta$ and $r\le 2a$) are given at Eqn. \ref{eqn-mobility-tensors-integral-odd} and its corresponding dimensionless explicit expressions are shown below\\

\begin{subequations}
\begin{align}
    \hat{M}\indices{^o_{UF}^{\alpha\beta}_{ij}}&=\chi\left(\frac{2}{3}-\frac{1}{8}r\right)\varepsilon_{ij3}-\frac{\chi}{8}r \varepsilon_{ijk} \frac{r_kr_3}{r^2}\\
    \hat{M}\indices{^o_{UT}^{\alpha\beta}_{ij}}&=-\chi\left(\frac{2}{5}-\frac{5}{32}r\right)\delta_{ij}r_3 + \chi \left(\frac{1}{10}-\frac{1}{32}r\right) \left(\delta_{i3}r_j + r_i\delta_{j3}\right) - \frac{\chi}{32r}r_i r_j r_3\\
    \hat{M}\indices{^o_{US}^{\alpha\beta}_{ijk}}&=\chi\left(\frac{1}{10} - \frac{1}{32}r\right)\left[\left(\varepsilon_{ij3}r_k + \varepsilon_{ik3}r_j\right)+\left(\varepsilon_{ijm}\delta_{k3}+\varepsilon_{ikm}\delta_{j3}\right)r_m\right]\\
    &\notag-\frac{\chi}{32r}\left(\varepsilon_{ijm}r_k+\varepsilon_{ikm}r_j\right)r_mr_3\\
    \hat{M}\indices{^o_{\Omega T}^{\alpha\beta}_{ij}}&=\frac{\chi}{2}\left(1-\frac{9}{16}r+\frac{1}{32}r^3\right)\varepsilon_{ij3}-\frac{3\chi}{2}\left(\frac{3}{16}r - \frac{1}{32}r^3\right)\varepsilon_{ijl}\frac{r_lr_3}{r^2}\\
    \hat{M}\indices{^o_{\Omega S}^{\alpha\beta}_{ijk}}&=-\chi\left(\frac{1}{5} - \frac{3}{32}r + \frac{1}{256} r^3\right)\delta_{i3}\delta_{kj}\\
    &\notag+\frac{3\chi}{2}\left(\frac{1}{5}-\frac{1}{8}r + \frac{1}{128}r^3\right)\left(\delta_{ij}\delta_{k3}+\delta_{ik}\delta_{j3}\right)\\
    &\notag-\frac{3\chi}{256}\left(8r+r^3\right)\frac{r_ir_jr_kr_3}{r^4}-\frac{3\chi}{256}\left(16r-3r^3\right)\left(\delta_{ik}\frac{r_jr_3}{r^2}+\delta_{ij}\frac{r_kr_3}{r^2}\right)\\
    &\notag+\frac{3\chi}{256}\left(8r-r^3\right)\left(\delta_{i3}\frac{r_kr_j}{r^2}+\delta_{k3}\frac{r_ir_j}{r^2}+\delta_{j3}\frac{r_ir_k}{r^2}+\delta_{jk}\frac{r_ir_3}{r^2}\right)\\
    \hat{M}\indices{^o_{E S}^{\alpha\beta}_{ijkl}}
    &=-\chi\left(-\frac{1}{10}+\frac{3}{64}r-\frac{1}{512}r^3\right)\left[\left(\varepsilon_{ik3}\delta_{lj}+\varepsilon_{il3}\delta_{kj}\right) + \left(\varepsilon_{jk3}\delta_{li}+\varepsilon_{jl3}\delta_{ki}\right)\right]\\
    &\notag+\frac{3\chi}{512}\left(8r+r^3\right)\left[\left(\varepsilon_{ikn}\frac{r_l}{r} + \varepsilon_{iln}\frac{r_k}{r}\right)\frac{r_j}{r}+\left(\varepsilon_{jkn}\frac{r_l}{r} + \varepsilon_{jln}\frac{r_k}{r}\right)\frac{r_i}{r}\right]\frac{r_nr_3}{r^2}\\
    &\notag-\frac{3\chi}{512}\left(8r-r^3\right)\left[\left(\varepsilon_{ik3}\frac{r_l}{r} + \varepsilon_{il3}\frac{r_k}{r}\right)\frac{r_j}{r} + \left(\varepsilon_{jk3}\frac{r_l}{r} + \varepsilon_{jl3}\frac{r_k}{r}\right)\frac{r_i}{r} \right.\\
    &\notag+\varepsilon_{ikn}\frac{r_n}{r}\left(\delta_{l3}\frac{r_j}{r} + \delta_{lj}\frac{r_3}{r} + \delta_{j3}\frac{r_l}{r}\right)+\varepsilon_{iln}\frac{r_n}{r}\left(\delta_{k3}\frac{r_j}{r} + \delta_{kj}\frac{r_3}{r} + \delta_{j3}\frac{r_k}{r}\right)\\
    &\notag+\left.\varepsilon_{jkn}\frac{r_n}{r}\left(\delta_{l3}\frac{r_i}{r} + \delta_{li}\frac{r_3}{r} + \delta_{i3}\frac{r_l}{r}\right)+\varepsilon_{jln}\frac{r_n}{r}\left(\delta_{k3}\frac{r_i}{r} + \delta_{ki}\frac{r_3}{r} + \delta_{i3}\frac{r_k}{r}\right)\right]
\end{align}
\end{subequations}

\subsection{Symmetry relations}\label{appx-sd-symmetry-odd}
A complete list of symmetry relations of odd mobility tensors between component indices ($i,j,k,l$), pairs indices ($\alpha,\beta$) and coupling indices ($X,Y$) ($X=U$, $\Omega$ or $E$ and $Y=F$, $T$ or $S$) is compiled below.
\subsubsection{Components indices}\label{appx-sd-symmetry-odd-space}
\begin{subequations}
    \begin{align}
    \hat{M}\indices{^o_{UF}^{\alpha\beta}_{ij}}&=-\hat{M}\indices{^o_{UF}^{\alpha\beta}_{ji}}\\
    \hat{M}\indices{^o_{UT}^{\alpha\beta}_{ij}}&=\hat{M}\indices{^o_{UT}^{\alpha\beta}_{ji}}\\
    \hat{M}\indices{^o_{US}^{\alpha\beta}_{ijk}}&=-\hat{M}\indices{^o_{US}^{\alpha\beta}_{ikj}}\\
    \hat{M}\indices{^o_{\Omega T}^{\alpha\beta}_{ij}}&=-\hat{M}\indices{^o_{\Omega T}^{\alpha\beta}_{ji}}\\
    \hat{M}\indices{^o_{\Omega S}^{\alpha\beta}_{ijk}}&=\hat{M}\indices{^o_{\Omega S}^{\alpha\beta}_{ikj}}\\
    \hat{M}\indices{^o_{E S}^{\alpha\beta}_{ijkl}}&=\hat{M}\indices{^o_{E S}^{\alpha\beta}_{ijlk}}=\hat{M}\indices{^o_{E S}^{\alpha\beta}_{jikl}}
    \end{align}
\end{subequations}

\subsubsection{Pairs indices}\label{appx-sd-symmetry-odd-parity}
\begin{subequations}
    \begin{align}
    \hat{M}\indices{^o_{UF}^{\alpha\beta}_{ij}}&=\hat{M}\indices{^o_{UF}^{\beta\alpha}_{ij}}\\
    \hat{M}\indices{^o_{UT}^{\alpha\beta}_{ij}}&=-\hat{M}\indices{^o_{UT}^{\beta\alpha}_{ij}}\\
    \hat{M}\indices{^o_{US}^{\alpha\beta}_{ijk}}&=-\hat{M}\indices{^o_{US}^{\beta\alpha}_{ijk}}\\
    \hat{M}\indices{^o_{\Omega T}^{\alpha\beta}_{ij}}&=\hat{M}\indices{^o_{\Omega T}^{\beta\alpha}_{ij}}\\
    \hat{M}\indices{^o_{\Omega S}^{\alpha\beta}_{ijk}}&=\hat{M}\indices{^o_{\Omega S}^{\beta\alpha}_{ijk}}\\
    \hat{M}\indices{^o_{E S}^{\alpha\beta}_{ijkl}}&=\hat{M}\indices{^o_{E S}^{\beta\alpha}_{ijkl}}
    \end{align}
\end{subequations}
\subsubsection{Generalized Lorentz reciprocal theorem}\label{appx-sd-symmetry-odd-time}
Based on the generalized Lorentz reciprocal theorem, six symmetry relations can be derived by applying it to different modes of motion.

\begin{subequations}
    \begin{align}
    \hat{M}\indices{^o_{UF}^{\alpha\beta}_{ij}}&=-\hat{M}\indices{^o_{UF}^{\beta\alpha}_{ji}}\\
    \hat{M}\indices{^o_{\Omega T}^{\alpha\beta}_{ij}}&=-\hat{M}\indices{^o_{\Omega T}^{\beta\alpha}_{ji}}\\
    \hat{M}\indices{^o_{E S}^{\alpha\beta}_{ijkl}}&=-\hat{M}\indices{^o_{E S}^{\beta\alpha}_{klij}}\\
    \hat{M}\indices{^o_{\Omega F}^{\alpha\beta}_{ij}}&=-\hat{M}\indices{^o_{UT}^{\beta\alpha}_{ji}}\\
    \hat{M}\indices{^o_{EF}^{\alpha\beta}_{ijk}}&=-\hat{M}\indices{^o_{US}^{\beta\alpha}_{kij}}\\
    \hat{M}\indices{^o_{ET}^{\alpha\beta}_{ijk}}&=-\hat{M}\indices{^o_{\Omega S}^{\beta\alpha}_{kij}}
    \end{align}
\end{subequations}

\subsubsection{Definition of the Fax\'en laws}\label{appx-sd-symmetry-odd-def}
\begin{subequations}
    \begin{align}
    \hat{M}\indices{^o_{\Omega F}^{\alpha\beta}_{ij}}&=\hat{M}\indices{^o_{U T}^{\alpha\beta}_{ij}}\\
    \hat{M}\indices{^o_{E F}^{\alpha\beta}_{ijk}}&=-\hat{M}\indices{^o_{U S}^{\alpha\beta}_{kij}}\\
    \hat{M}\indices{^o_{E T}^{\alpha\beta}_{ijk}}&=-\hat{M}\indices{^o_{\Omega S}^{\alpha\beta}_{kij}}
    \end{align}
\end{subequations}


\bibliographystyle{unsrt}
\bibliography{Literatures}

\end{document}